\documentclass[graybox]{svmult}
\usepackage{mathptmx}       
\usepackage{helvet}         
\usepackage{makeidx}         
\usepackage{graphicx}        
\usepackage{multicol}        
\makeindex             
\begin{document}
\title{Quantum Hydrodynamics}
\author{Shabbir A. Khan and Michael Bonitz}
\institute{Shabbir A. Khan \at National Centre for Physics,
Quaid-i-Azam University Campus, Islamabad 45320, Pakistan;
\\email: {sakhan@ncp.edu.pk} \and Michael Bonitz \at 
Institut f\"ur Theoretische Physik und Astrophysik,
Christian-Albrechts Universit\"at zu Kiel, 24098 Kiel, Germany;
\\email: {bonitz@physik.uni-kiel.de}}

\maketitle


\abstract{Quantum plasma physics is a rapidly evolving research
field with a very  inter-disciplinary scope of potential
applications, ranging from nano-scale science in condensed matter to
the vast scales of astrophysical objects. The theoretical
description of quantum plasmas relies on various approaches,
microscopic or macroscopic, some of which have obvious relation to
classical plasma models. The appropriate model should, in principle,
incorporate the quantum mechanical effects such as diffraction, spin
statistics and correlations, operative on the relevant scales.
However, first-principle approaches such as quantum Monte Carlo and
density functional theory or quantum-statistical methods such as
quantum kinetic theory or non-equilibrium Green's functions require
substantial theoretical and computational efforts. Therefore, for
selected problems, alternative simpler methods have been put
forward. In particular, the collective behavior of many-body systems
is usually described within a self-consistent scheme of particles
and fields on the mean-field level. In classical plasmas, further
simplifications are achieved by a transition to hydrodynamic
equations.
Similar fluid-type descriptions for quantum plasmas have been
proposed and widely used in the recent decade. This chapter is
devoted to an overview of the main concepts of quantum hydrodynamics
(QHD),\index{quantum hydrodynamics}\index{QHD \textit{see} quantum
hydrodynamics} thereby critically analyzing its validity range and
its main limitations. Furthermore, the results of the linearized QHD
in unmagnetized and magnetized plasmas and a few nonlinear solutions
are examined with illustrations. The basic concepts and formulation
of particle-particle interactions are also reviewed at the end,
indicating their possible consequences in quantum many-body
problems.}

\index{quantum plasma|(}
\section{Introduction}

Conventional plasmas found naturally in the visible universe (e.g.,
the Sun's environment, interplanetary and intergalactic media, etc.)
or created in the laboratory (e.g., discharge experiments, etc.) are
ionized gases in which the charged particles (electrons and
different ions) move under the influence of long-range
electromagnetic forces. Although, the individual particles obey the
laws of quantum mechanics, the wave nature of the particles has
practically no effect on the collective motion in the case of
classical plasmas, due to the large inter-particle distances, and
the plasma can adequately be described by classical dynamical laws
in the framework of Newtonian mechanics and Maxwell-Boltzmann (MB)
statistics.

It has been known since long ago that the conduction electrons in metals
 behave very similar to gaseous plasmas and can be well treated as an electron gas.
Similarly, electrons in semiconductors excited across the band gap
behave very similarly to a plasma of electrons (in the conduction
band) and holes (missing electrons in the valence band). This
electron-hole plasma is very similar to classical two-component
plasmas. However, there is a basic difference: the relevant
statistics changes from MB to Fermi-Dirac (FD), applicable to
identical quantum particles with half-integer spin whose
distribution is restricted by the Pauli exclusion principle. The
quantum electron gas in metals is globally neutralized by the
lattice ions whose properties are governed by various control
parameters (see below, for more details, see \cite{mb1}). Recent
developments in ultrafast spectroscopic techniques have made it
possible to monitor the collective behavior of the quantum electron
gas confined in nanomaterials (nanotubes, metal clusters,
nanoparticles, etc.) at the femtosecond scale. The collective
electron oscillations which are principally governed by plasma
effects lead to fascinating paradigm of plasmonics -- a research
field currently under way at a breathtaking pace
\cite{sm,ha,mb2,mb3}. In semiconductors, even though the electron
density is much lower than in metals, the ongoing miniaturization in
nanotechnology applications has made the spatial variations of the
doping profiles comparable to the de Broglie wavelength of the
electrons. This indicates the central role of typical quantum
effects, such as tunneling, on the behavior of future electronic
components. \index{electron}\index{ion}

Other realizations of quantum plasmas are obtained in high-density
matter. By using various dynamic and static compression techniques
(diamond anvils, gas guns, and so on), or high energy sources
(intense lasers, ion beams), dense plasma conditions with densities
of the order of $10^{23}{\rm cm}^{-3} \dots 10^{25}{\rm cm}^{-3}$\
have been achieved in the laboratory \cite{mb1}. At the initial
stage of compression, the temperature is moderate, and degenerate
electrons are expected. Significance of such experiments can be seen
in the warm dense matter (WDM) physics -- an area getting increasing
attention due to its importance for inertial confinement fusion
(ICF). The advent of superintense lasers (in the petawatt range and
beyond) provides tools for light matter-interactions expecting the
creation of overdense plasmas in the laboratory reaching an electron
number density up to $\sim 10^{26}{\rm cm}^{-3}$ \cite{sg1,cr}. Such
developments open the door to new avenues, making it possible to
understand the physics underlying various phenomena, experimentally
probing the quantum plasma regimes and, ultimately exploiting such
states for applications.
\index{warm dense matter}

Ionized quantum matter is found naturally in dense astrophysical
objects such as stellar cores, white and brown dwarfs, neutron
stars, and interior of giant planets (e.g., Jovian planets) in the
solar system \cite{chabrier}. The electrons in these systems
constitute a degenerate plasma that often is under extreme
conditions of density. Thereby, the electrons may be
non-relativistic or relativistic, depending upon the ratio of the
Fermi energy to the rest energy of an electron. At extremely high
densities, exceeding the nuclear density, $n\left( \sim 10^{39}{\rm
cm}^{-3}\right)$, nuclei break apart (Mott-like transition) giving
rise to a dense system of protons and neutrons. At still higher
density even protons and neutrons break up, turning into the exotic
quark-gluon plasma (QGP), a very special kind of quantum plasma
where the particles interact via a (color) Coulomb potential. Such
plasmas are believed to having existed immediately after the Big
Bang \cite{mt}, and seen in Relativistic Heavy Ion Collider (RHIC)
and Large Hadron Collider (LHC) experiments. For an overview on the
density-temperature range, see Fig.~\ref{fig:1}, for a more detailed
introduction to quantum plasmas, see Ref.~\cite{mb1}.

{\bf Basic parameters.} \index{coupling parameter} In many-particle
quantum systems, the mean particle distance $\bar{r_{i}} = [3
n_{i}/4\pi]^{-1/3}$\ of species {\textit {i}} is comparable to or
smaller than the de Broglie wavelength associated with the particle,
$\Lambda_{B_{i}} = h/\sqrt{2\pi m_{i}k_{B}T_{i}}$, where $m_{i}$\ is
particle mass and $T_{i}$ the temperature, leading to an overlap of
the wave functions of spatially extended mutually penetrating
quantum particles, and the quantum degeneracy parameter exceeds
unity, $\chi_{i} = n_{i}\Lambda_{B_{i}}^{3} \geq1$. For classical
systems, one can define the Coulomb coupling parameter\index{quantum
coupling parameter}\index{quantum plasma!parameters}
\index{Brueckner parameter} as the ratio of the average interaction
energy $\left\langle
U_{ii}\right\rangle=(e_{i}^2/4\pi\epsilon)\left(1/\bar{r_{i}}\right)$
and the average kinetic (thermal) energy $\left\langle
K\right\rangle =E_{Ti}$\ $\left(\sim k_{B}T_{i}\right)$, i.e.,
$\Gamma_{i}=\left\vert\left\langle U_{ii}\right \rangle\right \vert
/E_{Ti}$, where $\varepsilon$ is the static background dielectric
constant. But, for sufficiently cold and dense plasmas which are
quantum degenerate (assuming fermions), i.e. $\chi_i>1$, the role of
kinetic energy is taken over by the Fermi energy;
$E_{Fi}=\hbar^{2}\left(3\pi^{2}n_{i}\right)^{2/3}/2m_{i}$.\index{Fermi
energy}
In a quantum plasma, the strength of particle correlations is
measured by the Brueckner parameter; $ r_{si}=\bar{r_{i}}/a_{Bi}$,
where $a_{Bi}=(\varepsilon/e_{i}^{2})(\hbar^{2}/m_{i})$\ is the
effective Bohr radius. It is easily verified that $r_{si} \propto
\Gamma_{qi} = \left\vert\left\langle
U_{ii}\right\rangle\right\vert/E_{Fi} = (\hbar
\omega_{pi}/E_{Fi})^2$, where $\Gamma_q$ is the quantum
generalization of the Coulomb coupling parameter \cite{vt}, and
$\omega_{pi}$ is the plasma frequency given by $\omega^2_{pi}=n_i
e_i^2/\varepsilon_{0}m_i$ for particles in vacuum (in three
dimensions).
The ideal behavior is recovered for $\chi_{i}\ll 1$, $
\Gamma_{i}\ll 1$, in classical and $\chi_{i}\gg 1$, $r_{si}\ll 1$,
in degenerate quantum plasmas. Both these limits correspond to a structureless gas-like system and
are simple to analyze theoretically. The quantum coupling parameter
shows the peculiar property of dense quantum systems: they become increasingly
 ideal with increasing density\footnote{Note that this is different from ultrarelativistic quantum plasmas where kinetic and
interaction energy have the same scaling with density.}. In
contrast, in classical systems, the strength of correlations
increases upon compression since the interaction energy increases as
$n^{1/3}$ but thermal energy remains constant. Various plasma
regimes are illustrated in the density-temperature phase diagram in
the following section.
\index{quantum plasma!parameters}

{\bf Additional parameters.} For completeness we list further
parameters of relevance in quantum plasmas. The Fermi energy is
related to a characteristic velocity, wave number and length scale:
the Fermi velocity, wave number and wave length, $v^2_F = 2 E_F/m $,
$k^2_F = 2m E_F /\hbar^2$ and $\lambda_F=2\pi/k_F$. Furthermore,
there exists a characteristic screening length - the Thomas-Fermi
length $\lambda_{TF}$ that replaces the Debye screening length of a
classical plasma, [$\lambda_D^2 = k_BT/(2\pi e^2 n)]$,
$\lambda_{TF}= v_F/(\sqrt{3}\omega_p)$. \index{Debye length}

{\bf Theoretical concepts for quantum plasmas.} In quantum plasmas,
strong inter-particle interactions at de-Broglie length scale impede
the use of conventional classical theoretical models. That's why the
early descriptions of the most immediate quantum plasma--the quantum
electron gas in metals employed different approaches based on
semiclassical or quantum mechanical methods including some
fundamental works of the pioneers of the field
\cite{db1,yk1,yk2,dp1,dp2,dp3}. To find the properties of quantum
plasmas obeying FD or Bose-Einstein (BE) statistics, the
$N$-particle Schr\"odinger equation is the key equation which
describes the evolution of a general pure quantum state arising from
some initial state whereas the dynamics of the system is governed by
the Hamiltonian. In addition, the solution has to be
anti-symmetrized for the case of fermions (symmetrized for bosons).
\index{anti-symmetrization}
\index{quantum plasma!simulations}

For quantum plasmas, usually the description in terms of mixed
states is more appropriate. Then, instead of the wave function, the
system is described by the density operator ${\hat \rho}$, and the von Neumann
equation is the central equation that governs the dynamics of ${\hat \rho}$. For many-particle problems, the
computational tools based on the (time-dependent) Hartree-Fock (HF,
TDHF) method derivable from various techniques \cite{mb1,gm}
provides a useful path which allows for a solution of the
many-particle Schr\"odinger equation in an approximate way,
accurately describing and simulating the quantum and spin effects at
weak coupling.

If coupling is strong quantum kinetic methods become very demanding.
For equilibrium properties a powerful tool is Quantum Monte Carlo,
for instance Path Integral Monte Carlo (PIMC) \cite{mb1,mb4,vf1}.
This method is a very successful first-principle approach avoiding
model assumptions, and is well suited for bosonic particles. At the
same time, for fermions,  it is limited to small systems, due to the
so-called fermion sign problem. Equilibrium properties of correlated
quantum systems can also be described by quantum molecular dynamics
(QMD) techniques which include, for instance, the Wigner function
QMD \cite{vf2}, or
 classical MD with quantum and spin effects included via
effective quantum potentials\textrm{\ }\cite{mb5}. For equilibrium
solutions, theories like the random phase approximation (RPA) and
quantum mechanical modeling by density functional theory (DFT)
\index{density functional theory}\index{DFT \textit{see} density
functional theory} \cite{ph,ws} are also very successful. The DFT
has a vast range of applicability from atoms, molecules, solids to
classical and quantum fluids, and is generalized to deal with many
different situations. We further mention ideas to map a quantum
system to an effective classical one due to Dharma wardana et al.
\cite{dharma,perrot-dharma} and Dufty et al.
\cite{dufty-dutta,dutta-dufty1,dutta-dufty2}.

The standard description of {\em non-equilibrium} quantum plasmas is
based on kinetic theory which involves density matrices or phase
space distribution functions of coordinates and momenta. The time
evolution of the distribution function is given by a quantum kinetic
equation (QKE)\index{quantum kinetic equation} which differs from
the corresponding classical kinetic equation in the appearance of
the explicit difference of arguments in the potentials,  creating a
nonlocal coupling due to finite spatial extension of quantum
particles \cite{mb6}. The self-consistent kinetic modeling is one of
the main tools in quantum plasma dynamics in which the notation of
phase-space is provided by the Wigner representation in terms of the
density matrix. The QKE is a numerically expensive,
integro-differential equation which provides the basis for various
semiclassical approximations and computational schemes. Furthermore,
for non-equilibrium processes, the widely applicable method of
non-equilibrium Green's functions (NEGF)\index{non-equilibrium Green Functions} has allowed to achieve significant
progress in the past few years \cite{mb6,dk1,mb7,mb8,mb9}. It can
successfully describe the ultrafast dynamics of many-particle
systems and allows for a self--consistent treatment of the
collective linear and nonlinear response of correlated Coulomb
electron systems and non-perturbative inclusion of external fields
and systematic many-body approximations via Feynman diagrams. It
also offers an alternative formulation and extension of the TDHF
method in terms of a generalized one-particle density matrix
$G\left( x,t; \bar{x},\bar{t}\right)-$ the Green's function, which
depends upon two space-time variables (in general, also including
the spin projection), whose evolution is governed by the
Kadanoff-Baym (KB) equations \cite{kb}. The KB method has been used
to investigate the dielectric properties of plasmon oscillation
spectrum with collision effects included in a systematic and
consistent way in a correlated electron gas \cite{nk}. The
developments in analytical and computational tools have led to a
number of excellent textbooks including \cite{mb1,mb4,mb6,dk2} and
review articles, for instance \cite{mb10,mb11,mb12,mb13,si,ps,sv}.
Finally, we mention that the progress is significant, however the
solution and detailed analysis of QKE or the full description of
many-particle wave functions have been major challenges from a
theoretical perspective for the last several decades.

Owing to the analytical complexity of the quantum kinetic approach,
drastically simplified macroscopic models (e.g., semiclassical
molecular dynamics or quantum hydrodynamics) have been frequently
adopted in recent years which can reproduce some of the salient
features of quantum plasmas, although not providing the same
detailed information which can be extracted from quantum kinetic
theory. However, one has a choice with the alternative of studying a
physical problem microscopically -- with inherent technical
difficulties -- or macroscopically with a less cluttered and simpler
approach which usually has a more restricted applicability range.

Out of the semiclassical approaches for theoretical description of
quantum systems, Bohmian quantum mechanics and quantum hydrodynamics
(QHD)\index{quantum hydrodynamics}\index{QHD \textit  {see} quantum
hydrodynamics} have been widely used. The former considers real
particles in the classical sense of having their configuration space
trajectories determined by the Newtonian mechanics with specific
positions and momenta (the so-called hidden variables). The latter
is a more general method applicable to both pure and mixed states of
quantum statistical systems. The QHD equations are usually obtained
by taking moments of the appropriate kinetic equation (e.g., the
Wigner function equation) in analogy with the moments of the
classical kinetic equation. This leads to the conservation laws for
particle number, momentum and energy in terms of macroscopic
variables by choosing some suitable closure scheme in an approximate
way (for details, see below).

Since the early introduction by Madelung \cite{em}, various versions
of QHD have been developed and applied to many-particle bosonic and
fermionic systems, some of them have been mentioned above. For
instance, the QHD equations have been developed to study the
dynamics of the quantum electron gas in metals and thin metal films
\cite{gm1,gm2,gm3,nc,fh1}. For electrons in metals, the typical
electron density $n_{0}\simeq 10^{23}cm^{-3}$ yields the quantum
coupling parameter $r_{s}$ of the order of unity which apparently
shows that the collisionless models are not applicable to the
metallic electrons. However, the $e-e$ collision rate (inverse of
the electron lifetime $\tau _{ee}$) is controlled by the process of
Pauli blocking \cite{gm2}. At room temperature, $\tau_{ee}\simeq
10^{-10}$s which is much larger than the typical collisionless time
scale $\tau_{p}$, the inverse of the electron plasma frequency,
i.e., $\tau_{p}=\omega _{p}^{-1}\simeq 10^{-16}$s. In addition, the
typical relaxation time scale $\tau _{r}\simeq 10^{-14}$s is also
larger than $\tau _{p}.$ Therefore, for time scales smaller than
$\tau _{ee},$ the electron collisions can be neglected and the
collisionless models are appropriate. This standard justification of
QHD, however, has to be considered with great caution as it assumes
that the electron gas is in thermodynamic equilibrium. For example,
laser excited metals with non-equilibrium carrier distributions may
have much larger e-e scattering rates, despite the Pauli blocking
mechanism.

Similarly, the hydrodynamics formulation is applicable to
semiconductors which provides useful explanation of resonant
tunneling processes and many ultrafast phenomena at ultrasmall
scales \cite{aj,cg}. The model has also been extended to plasmonics,
for instance, surface-plasmon dispersion \cite{ab}, plasmonic device
applications \cite{mm}, and so on. For low-temperature bosons (e.g.,
Bose-Einstein condensates (BECs) in trapped Bose gases), the
elementary excitations and related phenomena can be seen by
employing the Gross-Pitaevskii theory in the spirit of QHD
\cite{eg,lp}. The model has also been applied to high gain free
electron lasers \cite{as}, and dense astrophysical plasmas
\cite{ak}, with the possibility of the inclusion of effects like
relativity and magnetic fields. However, here as well one has to
carefully examine the applicability limits of QHD. Many of the
recent predictions of QHD have to be treated with great care as long
as no experimental verification is possible or tests against more
accurate kinetic approaches have not been made. This has to be
re-iterated since many of the QHD papers are neglecting these
applicability limits and do not provide the necessary tests of their
results, see below.

In this chapter, we review the main concepts and limitations of QHD
and its validity in various applications starting from the simple
case of the weakly coupled, non-relativistic plasma in the
electrostatic limit. Since the topic of QHD is not new some
obvious derivations are not included and the reader is referred to appropriate
references. We start the introduction to the
method with a brief note on the initial proposals (Sec. 1.2) and a
discussion of the main assumptions and applicability conditions
discussing electron and ion plasma waves within linearized QHD. In
addition, a brief overview on some nonlinear solutions of QHD as
well as results for a magnetized quantum plasma are included with
a focus on the relevant low-frequency modes. The basic concepts of
correlations and their implications in quantum plasmas are then
introduced in Sec. 1.3. Finally, we discuss some recent problems
that are related to an incorrect application of QHD to hydrogen bound states
and spin effects in dense quantum plasmas.
The intention of this chapter is to discuss
the concepts in a pedagogical manner giving the interested readers
recommendations for suitable additional references and text books
for a more detailed study.
\index{quantum plasma|)}

\section{Basics of Quantum Hydrodynamics}
The hydrodynamic formulation\index{quantum hydrodynamics} of systems
which demonstrate behavior implicit in quantum mechanical subsystems
is almost as old as the Schr\"odinger equation. It started in the
early days of quantum mechanics when Madelung proposed that the
Schr\"odinger equation for spinless one-electron problems can be
transformed into the form of hydrodynamics equations. By taking the
complex wave function of the form $\psi =\alpha e^{i\beta }$ with
time-dependent, real valued $\alpha $\ and $\beta ,$\ he derived the
continuity equation and Euler-like equation from the Schr\"odinger
equation. Later on, after a long pause, Bohm and others played a
major role in the further developments in this direction. This
so-called Madelung hydrodynamics is usually considered as a
precursor of the Bohmian mechanics--a quantum theory based on causal
interpretation in terms of hidden variables $\cite {db2,db3,db4}$\
in which the reinterpretation of the solution of the Schr\"odinger
equation and associated phenomena on the lines of classical dynamics
was proposed. This interpretation is also known as de Broglie-Bohm
theory due to the idea of the pilot-wave by L. de Broglie carried
forward by Bohm to its logical conclusion.

\subsection{The time-dependent Schr\"odinger equation}\label{ss:tdse}
We begin by writing down the $N$-particle Schr\"odinger
equation\index{Schr\"odinger equation}
\begin{eqnarray} \label{eq:tdse}
 i\hbar \frac{\partial \Psi_N}{\partial t} &=& {\hat H}_N \Psi_N,
\\
 {\hat H}_N &=& \sum_{i=1}^N \left\{-\frac{\hbar^2}{2m}\Delta_i +
 V(\mathbf{r}_{i})\right\},
\label{eq:h}
\end{eqnarray}
which is supplemented by the initial condition for $\Psi_0$ for the
wave function at $t=0$. In Eq.~(\ref{eq:h}), $\Delta_i$ denotes the
Laplace operator (second spatial derivative with respect to the
coordinate of particle ``i''). The first term in the sum represents
the kinetic energy of the particles and $V({\bf r}_i)$ the potential
energy, just like in classical mechanics. Note that in (\ref{eq:h})
we neglect the interaction between the particles. Similarly, we have
disregarded the spin variables which will be discussed later.

Now, according to the ideas of Madelung and Bohm, the solution of
Eq.~(\ref{eq:tdse}) for time-dependent (in general complex)
$N$-particle wave function\index{wave function} is constructed with
the ansatz
\begin{equation}
\Psi_N(\left\{\mathbf{r}_{i}\right\} ,t)=A(\left\{
\mathbf{r}_{i}\right\} ,t)
\exp\left[\frac{i}{\hbar}\,
S(\left\{\mathbf{r}_{i}\right\} ,t)\right] ,
\hspace{0.5cm} {\rm where} \quad
\left \{
\mathbf{r}_{i}\right\}=\left\{r_{1},r_{2},\cdot\cdot
\cdot,r_{N}\right\},
\label{1}
\end{equation}
with $A(\left\{\mathbf{r}_{i}\right\} ,t)$\ being the real amplitude
function and $S(\left\{\mathbf{r}_{i}\right\} ,t)$, the real phase.
The statistical distribution of the trajectories determined from
$\left \vert \Psi_N \right \vert ^{2}$\ gives the probability
density that the particles are located at the coordinates $\left \{
\mathbf{r}_{i}\right\}$ and thus the measurable physical features of
the quantum system just like in classical statistics.

Inserting the ansatz (\ref{1}) for the wave function \ into the time-dependent
Schr\"{o}din\-ger equation (\ref{eq:tdse}) and separating the real and imaginary
parts, one obtains the following coupled equations for the two functions $A$\
and $S$
\begin{equation}
\partial_{t}S+\sum_{i=1}^{N}\left[ \frac{1}{2m}\left(\mathbf{\nabla}_{
\mathbf{r}_{i}}S\right)^{2}+V\left(\mathbf{r}_{i}\right)+
Q(\left\{\mathbf{r}_{i}\right\},t)\right] =0, \label{2}
\end{equation}
and
\begin{equation}
\partial _{t}A+\frac{1}{2m}\sum_{i=1}^{N}\left[ 2\left( \mathbf{\nabla}_{
\mathbf{r}_{i}}S\right)\cdot \mathbf{\nabla
}_{\mathbf{r}_{i}}A+A\nabla_{\mathbf{r}_{i}}^{2}S\right] =0,
\label{3}
\end{equation}
where, in addition to the conventional potential energy related to
$V\left(\mathbf{r}_{i}\right)$ there arises an new term,
$Q(\left\{\mathbf{r}_{i}\right\},t)$, that can be understood as
effective \textit{quantum} potential (or Bohm potential)\index{Bohm
potential} which is absent in the corresponding classical
system\footnote{Note again that the interactions between the
particles are neglected, hence all energy contributions are of
single-particle type. We will discuss the role of interactions
later, in Sec.\ref{s:interactions}.}. Bohm noticed that
Eq.~(\ref{3}) describes a conservation law of the probability
density whereas equation (\ref{2}) has the form of the classical
Hamilton-Jacobi equation with the generating function $S(\left
\{\mathbf{r}_{i}\right\},t)$, with an additional term given by
\begin{equation}
Q(\left\{\mathbf{r}_{i}\right\} ,t)=\frac{\hbar
^{2}}{2m}\frac{\nabla_{\mathbf{r}_{i}}^{2}A}{A}. \label{4}
\end{equation}
The associated ``effective Hamilton function'' now contains a total potential that is the
sum of the external potential and the quantum potential (summed over all particles)
and depends on the dynamical variables $\left\{\mathbf{r}_{i}(t),
\mathbf{p}_{i}(t)\right\}$\ such that the quasi-trajectories may be found
from
\begin{equation}
m\mathbf{\dot{r}}_{i}=\nabla_{\mathbf{r}_{i}}S(\left\{\mathbf{r}
_{i}\right\},t),\label{5}
\end{equation}
to yield $\mathbf{r_{i}}=\mathbf{r}_{i}(t)$ with initial position
$\bf{r_{0}},$\ where
$\mathbf{p}_{i}=\mathbf{\nabla}_{\mathbf{r}_{i}}S(\left\{\mathbf{r}_{i}\right
\},t).$\
The quasi-trajectories evolve under the influence of classical and
quantum potentials just like in classical mechanics. The key
difference is that the initial state of the particles is, in
general, not given in a deterministic manner, but different initial
coordinates occur with a finite probability that is given by the
initial continuous wave function $\Psi_0$. A computational
implementation of this scheme then requires a suitable statistical
procedure: one has to consider an ensemble of trajectories which
start from different initial conditions that have probabilities
(statistical weights) according to $|\Psi_0(\{{\bf r}_i\})|^2$. The
resulting time-dependent wave function is then obtained according to
the ensemble average of the individual trajectories with the same
weights. The above introduced QHD-description by a well defined wave
function can be extended to more general systems within a mixed
state representation given by density matrix, for more details, see
$\cite {mb6,phol,rw}$.

Another important remark has to be made. In fact, for a
many-particle system (even in a pure state) the Schr\"odinger
equation (\ref{eq:tdse}) does not provide the correct quantum
result. Since quantum particles are either bosons or fermions which
differ by their symmetry (in particular, the spin) the $N$-particle
wave function has to be symmetric (anti-symmetric) for bosons
(fermions), i.e. we need to apply a proper (anti-)symmetrization
procedure, $\Psi_N \rightarrow \Psi_N^{A/S}$, see. e.g. \cite{mb6,
mb10, mb13}. This has the well-known effect that even noninteracting
quantum particles become correlated with each other (or
``entangled''). This is fully included in the quantum kinetic
methods and simulations that were discussed in the preceding section
but is rarely discussed when applying the QHD approach. We will
return to this problem below in Sec.~\ref{s:conclusion}.

\subsection{Quantum mixed state description. Wigner function}\label{ss:wigner}
In 1932, E. Wigner \cite{ew} suggested the phase-space formulation
of quantum mechanics, a representation by means of joint
distributions of probabilities (more precisely, the
quasi-probabilities) for coordinates and momenta in phase space
which has led to another route to QHD. Wigner's original interest
was to find quantum corrections to classical statistical theory
where the Boltzmann factors contain energies expressible as
functions of both coordinates and momenta. The Wigner
function\index{Wigner function} doesn't necessarily stay
non-negative in its evolution process for some regions of phase
space due to restrictions on the simultaneous measurements of
coordinate and momentum by the Heisenberg uncertainty principle.
Unlike the classical case, it can therefore not be interpreted as a
true probability density. However, it is real, normalizable to unity
and gives averages just like the classical statistical distribution
function. The Wigner formalism has attracted considerable attention
in various disciplines of physics, and has also been the subject of
a detailed theoretical analysis, in turn motivating the efforts to
formulate various versions of quantum hydrodynamics due to the
analogy with classical fluid systems. For more details on
quasi-probability distribution functions and Wigner function method,
see \cite{vt,mb6,jm,pc,mh,hl}.

The Wigner function is a function of phase space variables $\left(
\mathbf{r},\mathbf{v}\right) $ and time (in the following we will
use velocities instead of momenta). For simplicity, we consider the
one-dimensional problem for quantum statistical mixture of states
$\left\{\psi_{i}(x,t),p_{i}\right\},$ $i=1,2,...,K,$ where each wave
function $\psi_{i}(x,t)$ is assumed with a real non-negative
probability $p_{i}$ $\left(0\leq p_{i}\leq 1\right) $ satisfying the
normalization condition $\sum \limits_{i=1}^{K}p_{i}=1.$ The results
are straightforwardly generalized to higher dimensions. When
correlations are ignored, the many-particle wave function can be
written as product of one-particle functions. Although we will
follow this idea below because it is at the heart of the QHD
approach, one has to clearly realize that this neglects the spin
properties of quantum particles. For fermions (or bosons) the
$N$-particle wave function - even if interactions are neglected -
{\em is not the product of single-particle wave functions} but has
the form of a Slater determinant (permanent). Therefore, all QHD
results so far assume that the associated exchange corrections
(terms additional to the simple product form) are not important.
Note that there is no guarantee for that and, depending on the
problem studied, the results may be quantitatively or even
qualitatively wrong, in particular if spin effects are studied, see
Sec.~\ref{s:conclusion}.

In this way, the quantum mixture (ensemble) of single-particle wave
functions is now represented by the density matrix (this is the
coordinate representation of a more general quantity -- the density
operator ${\hat \rho}$ \cite{mb9})\index{density matrix}
\begin{equation}
\rho\left( x',x'',t\right)=\sum\limits_{i=1}^{K}p_{i}\,\psi _{i}\left(
x',t\right)\psi _{i}^{\ast}\left( x'',t\right),  \label{6}
\end{equation}
where $^{\prime }\ast ^{\prime }$ denotes the complex conjugate, and
the sum extends over all states contributing to the mixture. The
next step is to introduce center of mass and relative coordinates,
$x=(x'+x'')/2$ and $s=x'-x''$, respectively which allows to rewrite
the original coordinates appearing in Eq.~(\ref{6}) as $x'=x+s/2$
and $x''=x-s/2$. As in classical kinetic theory, these coordinates
have different meanings: $x$ is related to the position of a
particle whereas the distance $s$ is related to the internal
structure and is the Fourier adjoint of the momentum\footnote{This
is seen by considering a spatially homogeneous system. Then all
points $x$ are equivalent, and the dependence on $x$ drops out,
whereas the dependence on $s$ remains.}. Then, the Wigner function
$f_{W}\left( x,v,t\right)$ can be written as the Fourier transform
of (\ref{6})\index{Wigner Transform} leading to
\begin{equation}
f_{W}\left( x,v,t\right) =\frac{m}{2\pi \hbar }\sum
\limits_{i=1}^{K}p_{i}\int_{-\infty }^{\infty }ds\psi _{i}^{\ast
}\left( x+\frac{s}{2},t\right) \psi _{i}\left(
x-\frac{s}{2},t\right) e^{imvs/\hbar },  \label{7}
\end{equation}
where $m$ is the particle mass and $v=p/m$ is the velocity.

As a side remark we mention that for a general $N$-particle system
with exchange and correlations described by a statistical mixture we
also can compute the single-particle Wigner function.\index{Wigner
function} However, then the starting point is the ensemble of the
full (anti-)symmetrized\index{anti-symmetrization} $N-$ particle
ensemble wave function $\Psi _{Ni}^{S/A}(x_{1},x_{2},...,x_{N},t)$.
Then the single-particle Wigner function follows from integration
over the variable of particles $2, 3, \dots N$,
\begin{equation}
f^{S/A}_{W}\left( x_{1},v_{1},t\right) =\int
dx_{2}dv_{2},...,dx_{N}dv_{N}f^{S/A}_{W N}\left(
x_{1},v_{1},...,x_{N},v_{N},t\right),  \label{8}
\end{equation}
where the integrand contains the (anti-)symmetrized $N$-particle
Wigner function which is the Fourier transform of the
(anti-)symmetrized $N$-particle density matrix as in the
one-particle case above \cite{mb9}. Instead of the known equation of
motion for the $N$-particle density matrix -- the von Neumann
equation -- one can also derive a chain of coupled equations for the
one-particle, two-particle etc. functions exactly like in the
classical case. This hierarchy is nothing but the quantum
Bogoliubov-Born-Green-Kirkwood-Yvon (BBGKY) hierarchy\index{BBGKY
hierarchy} of equations \cite{mb9}. As in the classical case, this
hierarchy can only be solved in special cases. In general, one has
to resort to
 closure approximations. In most cases one expresses the
two-particle function as a functional of single-particle functions
motivated by physical information about the system. The problem is
drastically simplified when particle correlations are neglected such
that the two-particle Wigner function is approximated as
$f_{W2}^{S/A}\left( x_{1},v_{1},x_{2},v_{2},t\right) =
\Lambda^{S/A}\{f_{W}\left( x_{1},v_{1},t\right) f_{W}\left(
x_{2},v_{2},t\right)\} $, where $\Lambda^{S/A}$ is the
(anti-)symmetrization operator. If further, exchange (and spin)
effects are being neglected we can drop the superscript ``S/A'', and
the two-particle function is just the product of two one-particle
functions. This is nothing but the quantum Vlasov (or Hartree)
approximation which is commonly used to derive the QHD equations.
This approximation means that the quantum plasma is considered as an
ensemble of particles interacting through a mean field potential.

The equation of evolution\index{quantum Liouville equation} for the
one-particle Wigner function (\ref {7}) for a scalar potential $V$
included in (\ref{2}) is given by
\begin{eqnarray}
&& \frac{\partial f_{W}}{\partial t}+v\frac{\partial f_{W}}{\partial
x} \label{9}
\\\nonumber
&&-\frac{i m}{2\pi \hbar^{2}}\int\int dsd\bar{v}e^{im\left(
v-\bar{v}\right) s/\hbar}\left[ V^{\rm eff}\left( x+\frac{s}{2}\right)
-V^{\rm eff}\left(x-\frac{s}{2}\right)\right] f_{W}\left(x,\bar{v},t\right)
=0,
\end{eqnarray}
and is obtained from the Wigner transform (\ref{7}) of the von
Neumann equation (or quantum Liouville equation) for the
single-particle density matrix \cite{mb9}. Here, $V^{\rm
eff}=V+V^{ind}$ is the total self-consistent potential that contains
the mean field potential $V^{ind}$ (in the case of Coulomb
interaction it is given by the solution of Poisson's equation)
exactly like in the classical Vlasov equation\footnote{The
derivation of Eq.~(\ref{9}) follows from straightforward algebra and
can be found in many text books, e.g. \cite{mb6}, and will,
therefore, not be reproduced here.}.

The underlying idea is that the quantum transport can be
seeded into generalized kinetic equation in the spirit of the
Boltzmann equation, appropriately extended with terms that
represent quantum corrections. However, the resulting Wigner
kinetic equation gives rise to a nonlocal dependence of distribution
function on momentum (for details on quantum kinetic equation, see
\cite{mb6}). Due to the finite spatial extension of the quantum
particles, the value of the potential energy at one space point also
depends on the values of $V$ at all other points--a pure quantum effect.
The classical limit is recovered in the limit of vanishing
difference of arguments of the two potentials\footnote{Then the integral term becomes
$\frac{1}{m} \frac{d V^{\rm eff}(x)}{dx} \frac{\partial f_W}{\partial v}$, as in the
classical Vlasov equation.}.

\subsection{Moments of the Wigner function. Hydrodynamics. }\label{ss:hydro}
The QHD model \cite{gm2} can be obtained by taking the moments of
(\ref{9}). Since, for all hydrodynamic approaches, the j-th order
moment requires the knowledge of the j+1-th moment, an infinite
chain of equations is found which demands a suitable truncation
scheme. Generally, the lower-order moments are related to physically
relevant quantities such as the particle density, average velocity,
and pressure etc. For mixed states, the pressure tensor (in higher
dimensions) requires the second moment of the Wigner function
equation to couple to the third moment. The closure assumption
allows to establish a relationship between the electron pressure and
density demanding an appropriate equation of state. This is, in general, a subtle issue, however, if the
system is in thermodynamic equilibrium (or sufficiently close) the
known equilibrium results for the equation of state can be used.

Defining the macroscopic variables, i.e., density, mean velocity and
pressure in the usual way ($\sigma^2_u$ is the variance of the velocity):
\begin{eqnarray}
n\left( x,t\right)=\int f_{W}(x,v,t)dv,\hspace{0.3cm} u\left( x,t\right)
=\frac{1}{n}\int f_{W}(x,v,t)\,v\,dv, \label{10}\\
\frac{1}{m}p\left(x,t\right) = \sigma^2_u = \int f_{W}(x,v,t)\,v^{2}dv-n(x,t)u^{2}(x,t) \label{11}
\end{eqnarray}
and representing each single-particle orbital
\begin{equation}
\psi _{i}(x,t)=A_{i}(x,t)\exp (iS_{i}(x,t)/\hbar ),  \label{12}
\end{equation}
with real amplitude $A_{i}(x,t)$ and real phase $S_{i}(x,t)$, the
first two equations become
\begin{eqnarray}
\frac{\partial n}{\partial t}+\frac{\mathbf{\partial }\left(
nu\right) }{\partial x}
&=&0,  \label{13} \\
m\left( \frac{\partial }{\partial t}+u\frac{\mathbf{\partial
}}{\partial x}\right) u
&=& -\left( \frac{\mathbf{\partial }V^{\rm eff}}{\partial
x}+\frac{1}{n}\frac{\partial p}{\partial x}\right),  \label{14}
\end{eqnarray}
where $V^{\rm eff}=V+V^{ind}$ is the
single-particle potential\footnote{$V$ is in fact non-local, which
can be seen from Eq.~(\ref{9}), but this is neglected in the
hydrodynamic formulation.} and $p$ is the total scalar pressure
which consists of two terms, $p=$ $p^{c}+p^{q}$, that will be
discussed below.

In the following, we consider the simple case of electrons in an
electrostatic field where the ions are treated as a homogeneous
background. Then $V^{\rm eff}= e \varphi$ with $\varphi =
\varphi^{ext} + \varphi^{ind}$,  which leads from (\ref{14}) to
\begin{equation}
\left( \frac{\partial}{\partial t}+u\frac{\mathbf{\partial
}}{\partial x}\right) u=\frac{e}{m}\frac{\mathbf{\partial }\varphi
}{\partial x}-\frac{1}{mn}\frac{\partial p^{c}}{\partial
x}-\frac{1}{mn}\frac{\partial p^{q}}{\partial x},  \label{15}
\end{equation}
where the induced electrostatic potential $\varphi^{ind} $ obeys Poisson's
equation
\begin{equation}
\frac{\partial ^{2}\varphi^{ind} (x,t)}{\partial x^{2}}=\frac{e}{\epsilon
_{0}}\left( \int dvf_{W}(x,v,t)-n_{0}\right) ,  \label{16}
\end{equation}
with $\epsilon _{0}$ and $n_{0}$ being the vacuum dielectric
constant and uniform background ion density, respectively.

In deriving the above QHD equations for the mean density $n$ and
mean velocity $u$, Eq. (\ref{10}), it is now crucial to have a
prescription how to connect them with the number density
$n_{i}\left( x,t\right) $ and velocity $ u_{i}\left( x,t\right) $
for each individual orbital. The latter are defined from the wave
function (\ref{12}) of each individual orbital according to $
n_{i}\left( x,t\right) =\left \vert \psi _{i}(x,t)\right \vert
^{2}=A_{i}^{2}(x,t)$ and $u_{i}\left( x,t\right) =\partial
_{x}S_{i}(x,t)/m$, i.e. just like in the case that the system is in
a pure state $\psi_i$ (as in Madelung's theory). This connection
follows readily from the definition of the density matrix (\ref{6}).
Thus the average with the Wigner function can be expressed as an
ensemble average
\begin{equation}\label{eq:ensemble-av}
 \langle \dots \rangle = \sum_{i=1}^K p_i \dots
\end{equation}
where the contribution of each orbital enters with the weight $p_i$.
Thus, for the mean density we obtain
\begin{equation}\label{eq:n}
 n(x,t)=\int dv f_W(x,v,t) = \sum_{i=1}^K p_i \int dv f_{Wi}(x,v,t) = \langle n_i(x,t)\rangle,
\end{equation}
and for the mean velocity follows analogously
\begin{eqnarray}\nonumber
n(x,t) u(x,t) &=& \int dv \,v\, f_W(x,v,t) = \sum_{i=1}^K p_i \int dv\, v\, f_{Wi}(x,v,t) =
\\\label{eq:u}
&=&
\sum_{i=1}^K p_i \, n_i(x,t)u_i(x,t) = \langle n_i(x,t)u_i(x,t)\rangle.
\end{eqnarray}
Finally, we obtain for the pressure from Eq.~(\ref{11}) two contributions: the first is the
same as in classical hydrodynamics,
\begin{equation}
\frac{p^{c}(x,t)}{m} = \sigma^2_u = \sum_{i=1}^K p_i \left( \int dv v^2 f_{Wi}(x,v,t) -
n_i(x,t) u_{i}^{2}(x,t)\right) ,  \label{18}
\end{equation}
and is given by the dispersion of the velocities. The orbital densities and velocities
will be eliminated from this expression below by postulating a suitable equation of state, see Eq.~(\ref{eq:eos}).

In contrast to classical hydrodynamics, here appears a second contribution to the pressure that arises from the coordinate dependence of the
orbital amplitudes in Eq.~(\ref{12}),
\begin{equation}
p^{q}(x,t)=\frac{\hbar^{2}}{2m}\sum\limits_{i=1}^{K}p_{i}\left[\left(
\frac{\partial A_{i}}{\partial x}\right)^{2}-A_{i}\frac{\partial
^{2}A_{i}}{\partial x^{2}}\right] ,  \label{17}
\end{equation}
which has been called {\em quantum pressure}
\cite{gm2}.\index{quantum pressure}

Thus the mean values $n$, $u$, $p^c$ and $p^{q}$ can be computed if the wave functions
 of all orbitals and, hence, the density matrix (Wigner function) are known, which is in general a very
difficult task.
%
Instead one can try to get a closed set of hydrodynamic equations, by invoking an equation of
state [hydrodynamic closure relation] that relates $p^{c}$ and $p^{q}$ to the macroscopic density
$n\left( x,t\right)$, thereby eliminating the individual $n_i$. In Ref. \cite{gm2} a very simple
solution was proposed:
The authors assumed a particular statistical mixture of states  in which {\em all single-electron wave functions (orbitals) $\psi_i$ have identical amplitudes} that are allowed to be space-dependent, i.e., $A_{i}\left( x\right) =A\left(
x\right), \; i=1,\dots K $. At the same time the different $\psi_i$ are allowed to have different phases, $S_{i}$, that are related to
the mean orbital velocity $u_{i}$ through the relation
$mu_{i}=\partial S_{i}/\partial x$ whereas the $u_{i}$ are related to the
global mean velocity $u$ via relation (\ref{eq:u}). This condition
with the help of (\ref{7}) and (\ref {11}) gives the density
$n=A^{2}.$ Also, this can be understood as an assumption of
uncorrelated electrons where the spatial distribution of each
electron defined by the amplitude $A_{i}$ doesn't depend upon the
spatial distribution of the other electrons in the system \cite{vt}.
This is a key assumption of QHD for a many-fermion system, and we
will discuss and test it more in detail in Sec.~\ref{ss:examples}.

With this assumption, the relation for $p^{q}$ can be rewritten as
\begin{equation}
p^{q}(n)=\frac{\hbar^{2}}{2m}\left[\left(\frac{\partial}{\partial
x}\sqrt{n}\right)^{2}-\sqrt{n}\frac{\partial^{2}}{\partial
x^{2}}\sqrt{n}\right], \label{19}
\end{equation}
where the last term in (\ref{15}) turns out to be
\begin{equation}
\frac{1}{mn}\frac{\partial p^{q}}{\partial x}=-\frac{\hbar
^{2}}{2m^{2}}\frac{\partial}{\partial x}\left(
\frac{1}{\sqrt{n}}\frac{\partial^{2}\sqrt{n}}{\partial
x^{2}}\right). \label{20}
\end{equation}
When compared to classical fluid equations for electrostatic
plasmas, the main difference is the Bohm potential term
(\ref{20})\index{Bohm potential} which takes the role of an
additional pressure. It is not a true pressure in thermodynamic
sense since it involves no velocity averages. In contrast, it is
caused by the quantum kinetic energy (which is proportional to minus
the Laplacian of the wave function) having the effect of particle
spreading (quantum diffraction, tunneling) which is formally
equivalent to a positive pressure.

To relate $p^{c}$ with the macroscopic density, a useful choice is
the equation of state\index{quantum equation of state} for strongly
degenerate (D-dimensional) fermion system in thermodynamic
equilibrium (we restrict ourselves to zero-temperature)\footnote{The
frequently used notion `classical' for $p^{c}\left(n\right)$ is
somewhat misleading because it contains $\hbar$ through $v_{F}$.
However it is analogous in the sense of measurement of the velocity
dispersion. }
\begin{eqnarray}
p^{c}(x) &=& p^{D}_{F}(n_0)\cdot \left(\frac{n(x)}{n_{0}}\right)^{5/3},
\label{eq:eos}\\
p^{D}_{F}(n_0) &=& \frac{2}{D+2} n_0 E_F(n_0) = n_0 \frac{mv_{F}^{2}}{D+2},
\label{21}
\end{eqnarray}
where $v_{F}=\hbar k_{F}/m$ is the Fermi velocity defined via the
electron Fermi wave number,\index{Fermi wave number}
$k_{F}=\left(3\pi^{2}n\right)^{1/3}$. Here $E_F(n_0)$ is the Fermi
energy of electrons in a homogeneous system of density $n_0$, and
$p^D_F(n_0)$ is the associated Fermi pressure of an ideal Fermi gas
at zero temperature. Note that the relation (\ref{21}) between
pressure $p_F^D$ and energy density $n_0 E_F$ is exact for a
non-relativistic ideal Fermi gas at $T=0$. The expression
(\ref{eq:eos}), on the other hand, extends this result to an
inhomogeneous system via the local approximation\footnote{An
alternative choice is a cubic dependence on $n(x)/n_0$ which is
 motivated in Ref.~\cite{vt} by assuming an adiabatic equation of state.}.

Using the assumption of identical orbital amplitudes, allows to
reformulate the quantum hydrodynamic equations (\ref{13}) and
(\ref{15}) as a nonlinear Schr\"odinger (NLS) equation
\cite{gm2}\index{Schr\"odinger equation!nonlinear}
\begin{equation}
i\hbar \frac{\partial \Psi }{\partial t}=-\frac{\hbar
^{2}}{2m}\frac{\partial ^{2}\Psi }{\partial x^{2}}-e\varphi \Psi
+{\tilde V}\Psi,  \label{22}
\end{equation}
where an effective wave function $\Psi (x,t)=\sqrt{n\left(
x,t\right) }\exp (iS(x,t)/\hbar )$ is defined with $mu\left(
x,t\right) =\partial S/\partial x,$ $n\left( x,t\right) =\left \vert
\Psi \right \vert ^{2}$ and $ {\tilde
V}=\frac{mv_{F}^{2}}{3n_{0}^{2}}\left \vert \Psi \right \vert ^{4}$.
The nonlinear Schr\"odinger-Poisson system captures the nonlinear
interaction between the electron density fluctuations and the
electrostatic potential. The NLS equation is easily amenable to
numerical analysis and its generalization can describe the behavior
of bosonic systems as well.

A related approach to derive the hydrodynamics equations is based on
the Dawson (classical) multistream model \cite{jd} which is extended
to the quantum case \cite{fh2} by considering a statistical mixture
of $K$ pure states representing $K$ ``streams'' of particles each
characterized by the same velocity. Following the Hartree
representation (well known in condensed matter physics), the states
with wave functions $\psi_{i}\left( x,t\right) ,$ $i=1,...,K$ obey
$K$ independent Schr\"odinger equations that are coupled via the
electrostatic potential \footnote{We mention again, that this
neglects exchange. Also the picture of ``streams'' of particles with
the same velocity is -- strictly speaking -- not compatible with the
Pauli principle.}
\begin{equation}
i\hbar \frac{\partial \psi_{i}}{\partial t}=-\frac{\hbar
^{2}}{2m}\frac{\partial^{2}\psi_{i}}{\partial x^{2}}-e\varphi \psi
_{i}. \hspace{1.2cm} i=1,...,K \label{23}
\end{equation}
Introducing the Madelung representation of wave function (\ref{12})
in (\ref {23}) and separating the real and imaginary parts, it
reduces to the hydrodynamic equations, the continuity equation, and
an Euler-like equation given by
\begin{equation}
\left(\frac{\partial}{\partial t}+u_{i}\frac{\mathbf{\partial
}}{\partial x}\right)u_{i}=\frac{e}{m}\frac{\mathbf{\partial
}\varphi}{\partial x}+ \frac{\hbar^{2}}{2m^{2}}\frac{\partial
}{\partial x}\left(\frac{\partial^{2}\sqrt{n_{i}}/\partial
x^{2}}{\sqrt{n_{i}}}\right), \label{24}
\end{equation}
where $n_{i}=A_{i}^{2}$. Setting $\hbar =0,$ the classical Dawson
relation \cite{jd} is retrieved. Although the equation (\ref{24})
takes into account the quantum diffraction effects, its
limitations are the same as those of (\ref{15}) and are described in the
following section.
\subsection{Examples and test of the assumption $A_i(x,t)=A(x,t)$}\label{ss:examples}
The assumption that all orbitals have the same space and
time-dependent amplitude, $A_i(x,t)=A(x,t)$, is a key assumption of
QHD for a many-fermion system at zero temperature\footnote{It is
also an assumption for a single particle at finite $T.$}. It is,
therefore, important to verify it.  In fact, the single-particle
wave functions $\psi_i$\index{wave function} are often easily found,
so it is possible to determine the amplitudes $A_i$ and phases
$S_i$, explicitly. We will do this
for three typical examples where, for simplicity we consider the one-dimensional case.\\

{\bf I. Homogeneous free electron gas.} We consider $N$ particles in
a box of length $2L$ with $-L\le x \le L$. To model a macroscopic
system, periodic boundary conditions are implied, i.e.
$\psi_i(-L)=\psi_i(L)$, for all $i$. If necessary, in the end one
can take the limit $L\to \infty$ and $N \to \infty$, while
maintaining a constant density $n_0=N/2L=$const. The solutions of
the one-particle Schr\"odinger equation that satisfy these boundary
conditions are well known ($E_l=\hbar^2k_l^2/2m$):
\begin{eqnarray}
\psi_l(x,t)&=&\frac{1}{\sqrt{2L}} e^{-i(E_l t -k_lx)/\hbar}, \quad
k_l= l k_0, \quad k_0 = \pi/L, \quad l= \pm 1, \pm 2, \dots
\nonumber\\
A_l(x,t) &=& A(x,t) = \frac{1}{\sqrt{2L}} = {\rm const}, \quad
S_l(x,t) = - E_l t  + \hbar k_l x. \label{eq:psi-e-gas}
\end{eqnarray}
Evidently, the amplitudes of all orbitals are equal.\index{orbital}\index{Schr\"odinger equation}\\

{\bf II. Non-interacting electrons in a deep potential well.}
Consider now the situation that the electrons are confined to a box
of length $2L$ with $-L\le x \le L$, where the potential walls are
assumed infinitely high. Then the electrons cannot penetrate into
the regions $x < -L$ and $x > L$ which leads to the boundary
condition on the wave functions $\psi_l(-L,t)=\psi_l(L,t)=0$, for
all $l$ and $t$. Again, the solutions to this problem are well known
from basic quantum mechanics. As usually, the time-dependent
solution is $\psi_l(x,t) = \psi_l(x)e^{-iE_l t/\hbar}$, where the
stationary solution is ($k_0=\pi/2L$)
\begin{eqnarray}
\psi_l(x)&=&\frac{1}{\sqrt{L}} \cos{k_l x}, \qquad  l
= \pm 1, \pm 3, \cdots ,
\nonumber\\
\psi_l(x)&=&\frac{1}{\sqrt{L}} \sin{k_l x}, \qquad  l
= \pm 2, \pm 4, \cdots,
\nonumber\\
A_l(x,t) &=& \psi_l(x), \qquad\quad S_l(x,t) = -E_l t.
\label{eq:psi-box}
\end{eqnarray}
In this case, the amplitudes $A_l$ of the orbitals are time-independent,
but all completely different, in striking contrast of the main assumption of
QHD.

{\bf III. Non-interacting electrons in a harmonic oscillator
potential.}  The previous case was characterized by a discontinuous
change of the external potential in space. Now we consider the case
of a smooth potential that is a quadratic function of the
coordinate, $V(x)=m\omega^2 x^2/2$. Then, the wave function can
extend into the whole space where, due to particle number
conservation (normalization condition), it should vanish
sufficiently fast for $|x|\to \infty$. The solution is again well
known, and the stationary wave functions are given by the Hermite
polynomials $H_l$,
\begin{eqnarray}
\psi_l(x)&=&\frac{1}{\sqrt{2^l l!\sqrt{\pi}x_0 }} \,
H_l(u)e^{-\frac{u^2}{2}}, \qquad  u = \frac{x}{x_0}, \quad x_0 =
\sqrt{\frac{\hbar}{m\omega}}, \quad l=0,1,2, \dots \nonumber\\
A_l(x,t) &=& \psi_l(x), \qquad \quad S_l(x,t) = -E_l t.
\label{eq:psi-osci}
\end{eqnarray}
As in the second case, the amplitudes of the orbitals are time-independent, but all different.

It is easy to understand the origin of this behavior.
In case I the solutions are freely propagating waves described by complex wave functions, and the amplitudes are equal.
In contrast, cases II and III correspond to bound states, where the electron motion is spatially restricted.
Correspondingly, the stationary wave functions are
real\footnote{This is not universal. For example for Coulomb bound states, the wave function is complex (the angular part).}
and the phases $S_l$ are just determined by the time-dependent exponential whereas the amplitudes are all different.
This also affects the quasi-classical velocities given by $mu_l = dS_l/dx$.
In case I we obtain $u_l= \hbar k_l/m$, whereas in cases II and III $u_l=0$ since $S_l$ is
independent of $x$ for all $l$. This is due to the fact, that bound wave functions correspond to standing waves with zero mean momentum.

The most important conclusion of these examples is that the {\em key assumption of QHD is not fulfilled for spatially confined electrons}.
While it is fulfilled for an infinite (noninteracting and spinless) system, this case does never occur in a real plasma. Furthermore, when studying waves
in quantum plasmas we are interested in the behavior of electrons in the presence of an external potential well which gives
rise to spatial confinement effects. Therefore, the case of confined particles is of particular importance.

It turns out that while the condition of equal orbital amplitudes
\cite{vt,mh}, $A_l(x,t)=\sqrt{n(x,t)}$, for all $l$, is certainly
sufficient for the validity of the QHD equations (together with an
appropriate equation of state), it is -- most likely -- not
necessary. In other words, the QHD equations may also be satisfied
if the condition $A_l(x,t)=\sqrt{n(x,t)}$ is not satisfied -- which
is practically never fulfilled. What is necessary for the QHD
equations is that all ensemble averages can be replaced by mean
quantities, in particular
\begin{eqnarray}
\frac{2m}{\hbar^2} p^{q}(x,t)&=&\left\langle\left(
\frac{\partial A_{i}}{\partial x}\right)^{2}-A_{i}\frac{\partial
^{2}A_{i}}{\partial x^{2}}\right\rangle
\Rightarrow
\left(\frac{\partial}{\partial
x}\sqrt{n}\right)^{2}-\sqrt{n}\frac{\partial^{2}}{\partial
x^{2}}\sqrt{n},
\label{eq:pq-qhd}
\end{eqnarray}
where $\langle ... \rangle$ denotes an average over the orbitals
with their weights $p_i$. These replacements can be understood as an
averaging procedure. Furthermore, note that the terms on the left
are rapidly varying in space (at least for the examples II. and
III.) because the amplitudes $A_l$ are oscillating increasingly
rapidly with growing $l$. This is not the case for the terms on the
right, i.e. for the
mean density $n$. Therefore, these replacements imply a suitable
spatial average which -- in principle -- is consistent with the
concept of a hydrodynamic approach. For the example II, an average
over the spatial period $\lambda_l=2\pi/ k_l$ of the square of the
amplitude $A_l^2$ yields $1/2$, for all $l$. However, problems
remain: it is not clear how to systematically choose a {\em single
length scale} averaging over which would apply to {\em all} (or at
least most) orbitals. In some cases, the Thomas-Fermi length
$\lambda_{TF}$,\index{Thomas-Fermi length} (or a multiple of it) may
be the proper scale as was found e.g. in Ref. \cite{bonitz_comment}.
Then for all orbitals with $\lambda_l \le \lambda_{TF}$ an averaging
occurs with $A_l^2 \Rightarrow 1/2$. Furthermore, the validity
condition of the assumption (\ref{eq:pq-qhd}) remains open.

\subsection{Main assumptions and applicability conditions of QHD}\label{ss:limitations}
\index{quantum hydrodynamics!applicability conditions}
The set of equations (\ref{13}), (\ref{15}) and (\ref{16})
constitutes a reduced model whose validity rests on several
assumptions, thus imposing important limitations on the model, as
described below. In the following we assume that only the electrons
are quantum degenerate, so all these conditions apply to the
electron component. Generalizations to several quantum components
are straightforward.
\begin{enumerate}
\item[(i)] The plasma is ideal (weakly coupled) which means all types of interactions
(or collision effects) are much weaker than the quantum kinetic
energy i.e.
\begin{equation}
 r_{s} \ll 1, \qquad {\rm or, \quad equivalently},\quad \Gamma_q \ll 1.
\label{eq:weak_coup}
\end{equation}
\item[(ii)] The interaction of the particles is treated in mean field approximation and
described by the induced electrostatic potential. No electromagnetic and quantization effects are
considered.

\item[(iii)] The wave phase velocities (as well as the particle velocities)
are non-relativistic i.e., $ \omega/k < c $.

\item[(iv)] The resolvable length scales in QHD are large ($L >$ several $\bar{r}$) which
means the equations are applicable to the long wavelength limit only
i.e., $\lambda \gg \lambda_{TF},$ or alternatively  $k\ll\omega_{p}/v_{F}$. Length scales shorter than the Thomas-Fermi
screening length $\lambda_{TF}$ $\left( =v_{F}/\sqrt{3}\omega_{p}\right) $, obviously,
cannot be resolved. This gives rise to a small parameter\footnote{
The long wavelength assumption is also
evident from analogy with the classical case. The assumption of the
classical pressure $p^{c}=mn\left( \left \langle u_{i}^{2}\right
\rangle -\left \langle u_{i}\right \rangle^{2}\right) $ leads to the
equation of state, $ p^{c}=p^{c}(n)$ for dense degenerate electrons
at $k_{B}T\ll E_{F}.$ This in turn demands the condition $k\lambda
_{TF} \ll 1$ to describe the wavelengths within the QHD just like the
classical fluid condition $k\lambda_{D} \ll 1$ with $\lambda_{D}$
$\left( =k_{B}T/\sqrt{4\pi ne^{2}}\right) $ being the Debye length.}
,
\begin{equation}
\kappa \equiv k\lambda_{TF} \ll 1. \label{eq:small_lambda}
\end{equation}
%
Kinetic phenomena such as Landau damping cannot be described by QHD since they depend on the details
of the equilibrium Wigner function and have to be treated with kinetic
theory.
\item[(v)] The equation of state of an ideal Fermi gas at $T=0$ is used within the local
approximation. The extension to space-dependent density profiles is
done by introducing a factor $(n(x)/n_0)^{5/3}$, cf.
Eq.~(\ref{eq:eos}). This again imposes restrictions on the
resolvable length scales. In particular, rapidly varying density
profile (e.g. in the case of steep external potentials) cannot be
treated properly because the non-locality would give rise to large
gradient effects. The zero-temperature assumption requires that
$T\ll T_F$, otherwise there will be corrections from the Sommerfeld
expansion of the Fermi function, e.g. to the equation of state.
\item[(vi)] No energy transport equation is considered. This could be done by
taking the second order moment of the Wigner function equation.
\item[(vii)] The model is approximate via the Bohm potential.
The closed set of equations follows if the amplitudes $A_{i}(x,t)$
of all single-particle orbitals are equal (but not necessarily
constant). These orbitals can have different phases $S_{i}(x,t)$
which are related to the mean orbital velocity through
$mu_{i}=\partial S_{i}/\partial x$, and $u_{i}$ is related to the
global mean velocity $u$ defined in (\ref{10}). This in turn implies
the same macroscopic density $n\left( x\right) =A^{2}(x)$. A less
restrictive condition is given by Eq.~(\ref{eq:pq-qhd}), see the
examples and discussion in the previous section.
\item[(viii)] No spin effects are taken into account. However, the
inclusion of a magnetic field in QHD is straightforward by starting
from a quantum kinetic equation with an electromagnetic field included
(by any gauge), as is briefly discussed on the following pages.
\end{enumerate}

As is often the case, a physical model may be valid even beyond its
formal conditions of applicability. This is also sometimes the case
with QHD which may give reasonable results even beyond the
conditions listed above. However, there is no guarantee for this,
and a careful analysis of the relevant conditions should always be
performed.

\begin{figure}
\centering
\includegraphics [width=12cm]{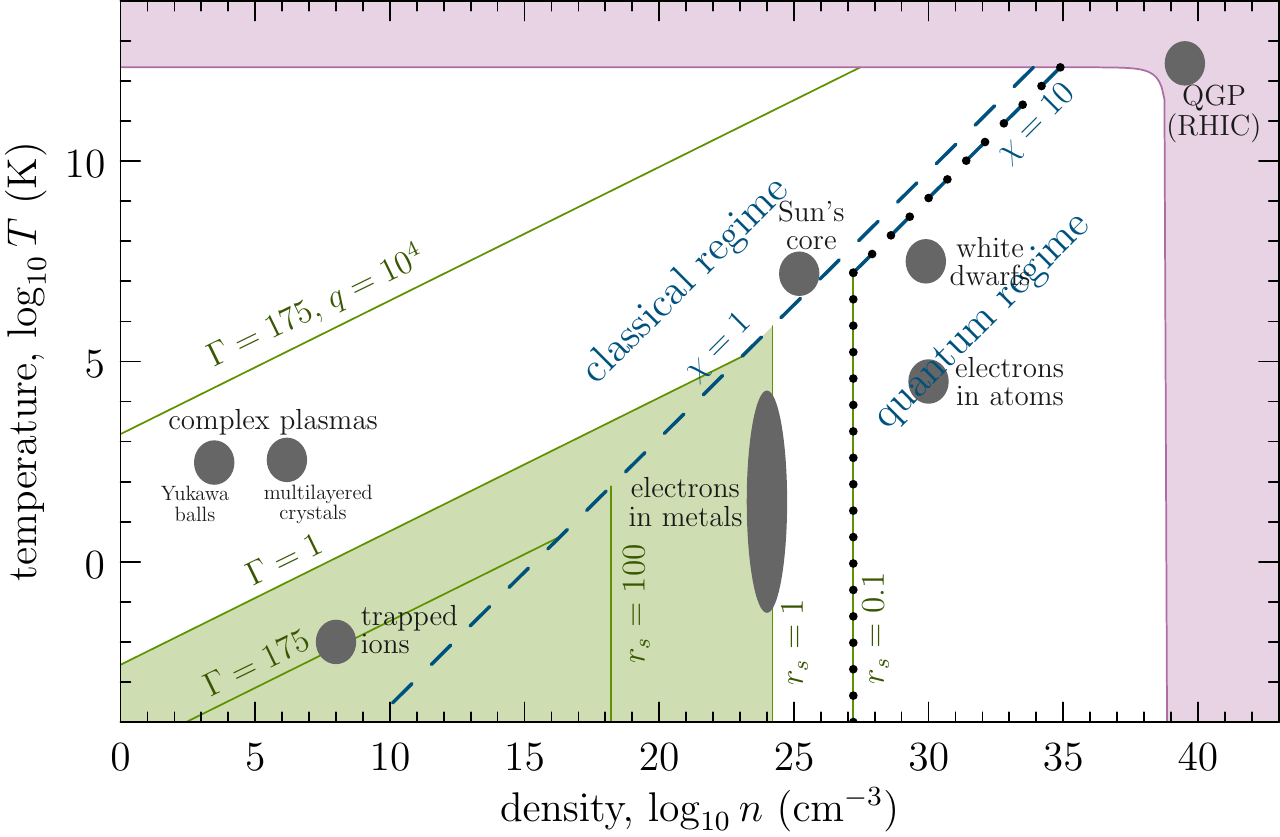}
\caption{Density-temperature phase diagram of a one-component
plasma, for example electrons in a neutralizing positive background.
The green triangle corresponds to strongly correlated electrons,
below (left from) the line $\Gamma=175$ ($r_s=100$) electrons form a
Wigner crystal. The green line $\Gamma=175$, $q=10^4$ corresponds to
crystallization in dusty plasmas containing particles with charge
$q=10,000e_0$. Quantum effects are relevant to the right of the
dashed line which is given by $\chi=1$. Some occurrences of quantum
plasmas are noted in the figure. QGP denotes the quark-gluon plasma
that (is thought to have) existed shortly after the Big Bang and
which also was produced at the relativistic heavy ion collider in
Brookhaven (RIC) and at the LHC at CERN. The restriction of QHD to
weak coupling, $r_s < 0.1$, corresponds to densities larger than
$10^{27}$cm$^{-3}$. The restriction to the ground state ($T=0$)
requires at least $\chi > 10$. Note that this makes applications of
QHD to electrons in metals or to warm dense matter very
questionable. This range is indicated by the dotted line.
\index{quantum plasma!phase diagram}\index{quantum hydrodynamics!applicability conditions}}
\label{fig:1}
\end{figure}

In what follows, we have shown in Fig. \ref{fig:1} the regions of
applicability of the QHD through the density-temperature phase diagram.\\
The properties of quantum electron gas can be measured with good
accuracy in hydrodynamic approximation and provide an ease to search
and analyze the linear waves and instabilities which gives insight
of the main role of quantum effects.
\subsection{Linearized QHD: Linear waves of quantum plasmas}
We begin the applications of QHD by considering the linear response
of electrons in a quantum plasma to a weak external excitation. Then
the QHD equations\index{quantum hydrodynamics} can be linearized
allowing to compute a dielectric function from which the plasmon
spectrum is straightforwardly obtained. The problem to study first
is electron plasma oscillations\index{electron plasma wave} where
many results exist against which the QHD result can be directly
tested.
\subsubsection{Electron plasma waves}
In order to apply the quantum hydrodynamics equations, consider a
zero-temperature fermion gas in one spatial dimension with the
pressure given by (\ref{21}) as follows
\begin{equation}
p^{c}(n)=\frac{mv_{F}^{2}}{3n_{0}^{2}}n^{3}. \label{25}
\end{equation}
The electron dynamics are governed by equations (\ref{13}),
(\ref{15})--(\ref{16}) whereas the ions are considered immobile,
forming a neutralizing background. Linearizing the equations around
the homogenous equilibrium; $n=n_{0}, u=0$ and $\varphi =0$, and
Fourier analyzing as usual with small fluctuating quantities, $n_1, u_1, \phi_1$, expressed as
$ \exp \left[ i\left(kx-\omega t\right)\right] ,$ one obtains the
dispersion relation
\begin{equation}
\omega^{2}=\omega_{p}^{2}+k^{2}v_{F}^{2}+\frac{\hbar^{2}k^{4}}{4m^{2}}
= \omega_{p}^{2}+k^{2}v_{F}^{2}\left(1 + \frac{3}{16}\Gamma_q \kappa^2 \right),\label{26}
\end{equation}
where $\omega $ is the wave frequency and $k$ the wave number. This
relation is also derivable from the Poisson-NLS equations (\ref{16})
and (\ref{22}) in the linear limit. While it is nice to keep in the
dispersion relation (\ref{26}) the fourth order term in $k$ one has
to clearly remember the limitations of QHD, see above. Indeed, the
last equality in (\ref{26}) shows that the $k^4$ term is about three
orders of magnitude smaller (considering that $\Gamma_q \ll 1$ and
$\kappa \ll 1$) than the $k^2$ term and there is no justification to
retain it within QHD.

\index{Wigner-Poisson equations} We now want to compare (\ref{26})
with the result obtained from the Wigner-Poisson model (\ref{9}) and
(\ref{16}). Assuming that the potential in (\ref{9}) depends on one
coordinate only, say $x$, we approximate the equation up to $O\left(
\hbar^{2}\right) $ given by \cite{mb6}
\begin{equation}
\frac{\partial f_{W}}{\partial t}+v\frac{\partial f_{W}}{\partial
x}+\frac{e}{m}\frac{\partial \varphi }{\partial x}\frac{\partial
f_{W}}{\partial v}=\frac{e\hbar ^{2}}{24m^{3}}\frac{\partial
^{3}\varphi }{\partial x^{3}} \frac{\partial ^{3}f_{W}}{\partial
v^{3}}+O(\hbar^{4}). \label{27}
\end{equation}
The right-hand side of (\ref{27}) is due to the non-locality of the
potential in the equation for the Wigner function (\ref{9}). It is
now easy to see that, in the limit $\hbar \rightarrow 0,$ one
recovers the familiar Vlasov equation for a classical collisionless
plasma. The result can be found in perturbation theory and using a
Fourier decomposition of the perturbations. Considering the
contribution from a monochromatic perturbation proportional to $\exp
\left[ i\left( kx-\omega t\right) \right] ,$ i.e.,
\begin{eqnarray}
f_{W}\left( x,v,t\right) &=& f_{0}\left( v\right) +f_{1}\left( v\right)
\exp \left[ i\left(kx-\omega t\right)\right], \nonumber\\
\varphi(x,t) &=& \varphi_{1}\exp \left[ i\left(kx-\omega t\right)\right] ,
\label{28}
\end{eqnarray}
where $f_{1}$ and $\varphi _{1}$ are first order perturbed
quantities and $\left \vert f_{1}\right \vert\ll f_{0}$. It leads to
the dispersion relation $\epsilon\left(\omega,k\right) =0$, where
the dielectric function $\epsilon$ for the Wigner-Poisson system\index{dielectric function}
reads\footnote{Here we assume that $\omega$ contains an infinitely
small imaginary part in order to assure causality (Landau pole
integration). i.e. $\epsilon$ is understood as a retarded
quantity.},
\begin{equation}
\epsilon\left(\omega,k\right) =1-\frac{m\omega _{p}^{2}}{n_{0}\hbar
k^{2}}\int\frac{f_{0}\left(v+\hbar k/2m\right) -f_{0}\left(v-\hbar
k/2m\right)} {kv-\omega }dv. \label{29}
\end{equation}
With a suitable change of variables\footnote{i.e. changing the
integration variable according to $v \rightarrow v \mp \hbar k/2m$)
and bringing both terms to a common denominator.}, the dispersion
relation for high frequency electron plasma oscillations becomes
\begin{equation}
\epsilon\left(\omega,k\right) =1-\frac{\omega_{p}^{2}}{n_{0}}\int
\frac{f_{0}\left(v\right)}{\left(\omega -kv\right)^{2}-\left(\hbar
^{2}k^{4}\right)/4m^{2}}dv=0. \label{30}
\end{equation}
This is just the Lindhard dispersion relation \cite{jl} which is
well known in solid state physics. In the one-dimensional case, the
equilibrium Wigner function for a fully degenerate Fermi gas is
given by
\begin{eqnarray}
f_{0}\left( v\right)=\frac{n_{0}}{2v_{F}}\hspace{1.2cm}\left\vert
v\right \vert <v_{F},\\ =0,\hspace{1.6 cm}\left\vert v\right\vert
>v_{F}, \label{31}
\end{eqnarray}
which leads to \cite{mb6}
\begin{equation}
\epsilon\left(\omega,k\right) =1-\frac{m\omega _{p}^{2}}{2\hbar
k^{3}v_{F}}\ln \left \vert \frac{\omega^{2}-\left( kv_{F}-\hbar
k^{2}/2m\right)^{2}}{\omega^{2}-\left( kv_{F}+\hbar
k^{2}/2m\right)^{2}}\right\vert =0. \label{32}
\end{equation}
In the long wavelength limit, $kv_{F}\ll \omega,$ $\hbar k^{2}/2m\ll
\omega ,$ expansion of $\epsilon\left(\omega,k\right) $ in
(\ref{32}) gives (\ref{26}), the limit of the kinetic dispersion
relation for small wave numbers.
\begin{figure}
\centering
\includegraphics [scale=1.2]{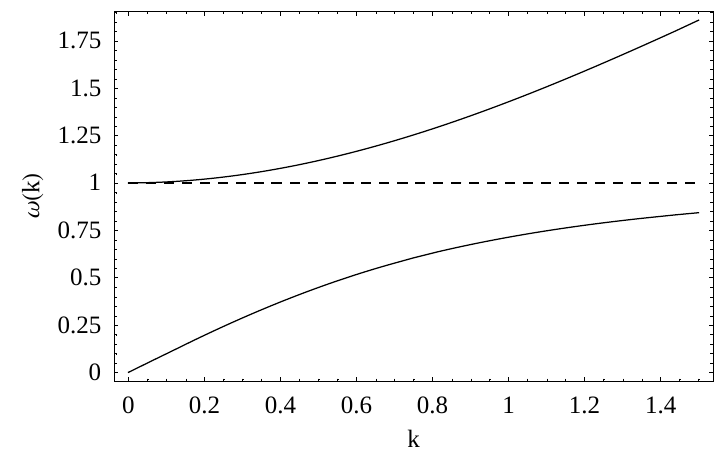}
\caption{The dispersion relations (\ref{26}) for electrons (upper
curve) and (\ref{51}) for ions (lower curve) are shown for the case
of  strong degeneracy, $(E_T \ll E_{F})$. The wave frequency for
electrons (ions) is normalized by $\omega_{p}(\Omega_{p})$ and the
wave number by $\omega_{p}/v_{F} (\Omega_{p}/c_{q})$.} \label{fig:2}
\end{figure}
Like for the one-dimensional case, the equilibrium function equals
zero for $v > v_F$. Only for smaller velocities $\left \vert
\mathbf{v}\right \vert <v_{F},$ the absolute value differs from 1D:
$f_{0}\left(\mathbf{v}\right) =\frac{n_{0}}{\pi v_{F}^{2}}$, for
$D=2$, and $f_{0}\left( \mathbf{v}\right)=\frac{n_{0}}{4\pi
v_{F}^{3}/3} $, for $D=3,$ reflecting the different normalization
conditions. So, combining the results for the different dimensions,
the dispersion relation\index{dispersion relation} for a fully
degenerate Fermi gas in the long wavelength limit takes the form
\begin{equation}
\omega^{2}=\omega _{p}^{2}+\left(\frac{3}{D+2}\right)
k^{2}v^2_{F}+\frac{\hbar^{2}k^{4}}{4m^{2}}, \label{33}
\end{equation}
which should be reproduced by the hydrodynamic equations. Quantum
mechanical effects enter this result in two distinct ways: the first
is statistical in the sense that the equilibrium distribution is the
Fermi distribution, and the second is quantum dynamical [the last
term in (\ref{33})], arising from the energy associated with the
finite momentum transfer $\hbar k$ of an electron interacting with a
plasma oscillation\footnote{The quantum picture describes this as
scattering of an electron with a quantum particle -- the plasmon.
This momentum change appears in the arguments of the distribution
functions in Eq.~(\ref{29}). The classical limit is obtained from
formally letting $\hbar \to 0$, then the difference of distribution
functions turns into a derivative with respect to momentum, and one
recovers the classical Vlasov dielectric function.}.

We note that this type of dispersion of electron plasma oscillations
(the quantum Langmuir-like wave as shown in Fig. \ref{fig:2}) is not
new and has already been found by Klimontovich and Silin \cite{yk1}
by using the Wigner distribution function, and by Bohm and Pines
\cite{db1} by developing canonical transformations of the
Hamiltonian of the system of electrons interacting through the
electrostatic force.

\begin{figure}
\centering
\includegraphics [scale=1.2]{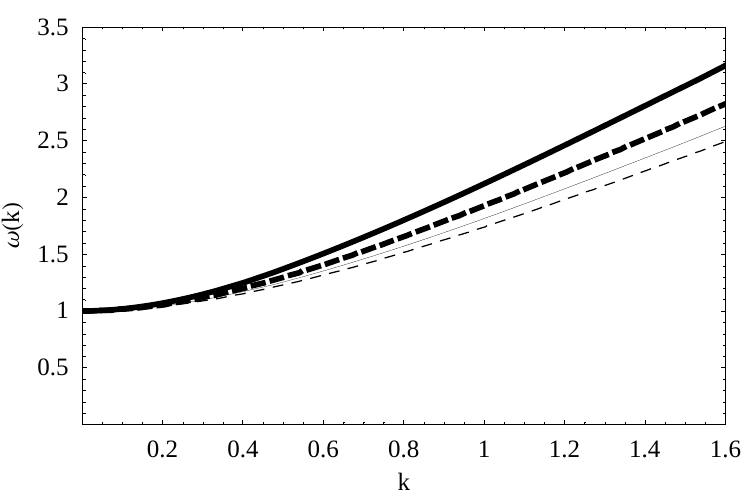}
\caption{The modified Bohm-Gross relation (\ref{34}) is shown with
variation in degeneracy parameter $\chi$=n$\Lambda_{B}^{3}$ for a fixed
quantum coupling parameter $r_{s}=0.1$. Thick (thick dashed) line
corresponds to $T=300 K$ ($T=500 K$) whereas thin (thin dashed )
line is for $T=700 K$ ($T=900 K$). The wave frequency is normalized
by $\omega_{p}$, and wave number by $\omega_{p}/v_{F}$,
respectively.} \label{fig:3}
\end{figure}

In analyzing the dispersive properties of quantum plasmas, the
coupling and degeneracy parameters play a key role in choosing an
appropriate model \cite{vg}. The quantum coupling parameter $r_{s}$
is a function of density only which shows that the higher is the
density of quantum particles, the weaker are the correlations in the
system. Some important parameters related to typical degenerate
laboratory and astrophysical plasmas are given in the following
tables.
\begin{table}
\caption{The parameters of a degenerate electron gas with number
densities of the order of metallic electrons with large but constant
degeneracy parameter ($\chi \gg 1$). $E_T=k_BT$. Since the electron gas is
moderately coupled ($r_s \sim 1$), this system is, strictly speaking, not accessible
to QHD.}
\label{tab:1}
\begin{tabular}{p{2.0cm}p{1.7cm}p{0.8cm}p{0.7cm}p{1.9cm}p{1.9cm}p{1.7cm}}
\hline\noalign{\smallskip} $n\left[10^{23} {\rm cm}^{-3}\right]$ &
$\bar{r}\left[10^{-9} {\rm cm}\right]$ & $r_{s}$ & $T\left(K\right)$
& $\omega_{p}\left[10^{16} s^{-1}\right]$ &
$E_{F}\left[10^{-11} {\rm erg}\right]$ &$E_{T}\left[10^{-14} {\rm erg}\right]$\\
\noalign{\smallskip}\svhline\noalign{\smallskip}
$3.0$ & $9.2$ & 1.10  & $300$ & $2.4$ & $2.6$ & $4.4$ \\
$3.7$ & $8.6$ & 1.02  & $350$ & $2.7$ & $3.0$ & $4.8$\\
$4.6$ & $8.0$ & 0.95  & $400$ & $3.0$ & $3.5$ & $5.5$ \\
$5.5$ & $7.6$ & 0.90  & $450$ & $3.3$ & $3.9$ & $6.2$\\
$6.5$ & $7.2$ & 0.84  & $500$ & $3.6$ & $4.3$ & $6.9$\\
\noalign{\smallskip}\hline\noalign{\smallskip}
\end{tabular}\\
\end{table}\\
\index{quantum plasma!parameters}
\begin{table}
\caption{Typical parameter range of a high-density degenerate electron gas
found in compact astrophysical systems such as dwarf stars. These systems are weakly coupled and
are well suited for a QHD description.}
\label{tab:2}
\begin{tabular}{p{1.0cm}p{1.9cm}p{1.9cm}p{1.9cm}p{1.9cm}p{1.9cm}}
\hline\noalign{\smallskip} $r_{s}$ & $n\left[{\rm cm}^{-3}\right]$ &
$\bar{r}\left[10^{-10} {\rm cm}\right]$  &
$E_{F} \left[{\rm erg}\right]$ & $v_{F}\left[{\rm cm/s}\right]$ & $\lambda_{TF}\left[10^{-8} {\rm cm}\right]$\\
\noalign{\smallskip}\svhline\noalign{\smallskip}
$0.10$ & $4.0\times10^{26}$ & 8.3  &  $3.1\times10^{-9}$ & $2.6\times10^{9}$ & $7.2$ \\
$0.08$ & $7.8\times10^{26}$ & 6.7  &  $4.9\times10^{-9}$ & $3.2\times10^{9}$ & $6.5$ \\
$0.06$ & $1.8\times10^{27}$ & 5.0  &  $8.8\times10^{-9}$ & $4.3\times10^{9}$ & $5.7$ \\
$0.04$ & $6.3\times10^{27}$ & 3.3  &  $1.9\times10^{-8}$ & $6.5\times10^{9}$ & $4.6$ \\
$0.02$ & $5.0\times10^{28}$ & 1.6  &  $7.9\times10^{-8}$ & $1.3\times10^{10}$ & $3.2$ \\
\noalign{\smallskip}\hline\noalign{\smallskip}
\end{tabular}
\end{table}
The significance of quantum dispersion effects of electron plasma
oscillations have been observed in solid-density plasmas. The plasma
compression experiments show that the plasmon frequency is a
sensitive measure of the electron density and the plasmon dispersion
relation includes the Fermi degeneracy effects. In these experiments the
temperature is finite and the above result is not applicable. On the other hand,
the plasmon dispersion of a classical plasma is well known. It starts from the
plasma frequency (for $k=0$) as well and then $\omega^2$ increases proportional to $k^2v_{th}^2$
[$v_{th}=\left(k_{B}T/m\right) ^{1/2}$ is the electron thermal
velocity] -- the so-called Bohm-Gross dispersion.\index{dispersion relation!Bohm-Gross} If, at finite $T$, quantum effects become relevant one
obtains the modified Bohm-Gross relation for small $k$ that contains quantum corrections \cite{sg1}:
\begin{equation}
\omega^{2}=\omega_{p}^{2}+3k^{2}v_{th}^{2}\left( 1+0.088n\Lambda
_{B}^{3}\right)+\frac{\hbar^{2}k^{4}}{4m^{2}}, \label{34}
\end{equation}
where $\Lambda_{B}$ is the thermal de Broglie wavelength. Since the
degeneracy parameter varies with temperature, the relation shows the
increase in wave dispersion with decreasing temperature [Fig.
\ref{fig:3}]. Such noticeable effects of fermion degeneracy
 in dense matter at relatively high temperature provide useful information about the
plasmon dispersion in future experiments, for details, see
\cite{sg1}.
\subsubsection{Dielectric tensor of a relativistic quantum plasma}
\index{dielectric tensor, relativistic} Let us turn to the full
dielectric tensor of a degenerate non-relativistic electron gas.
This tensor is well known since the 1950s \cite{yk1,jl} with various
generalizations to the fully relativistic quantum
regime.\index{quantum plasma!relativistic} The dielectric tensor of
an unpolarized isotropic electron--positron plasma \index{quantum
plasma!relativistic} can be written as \cite{dm1,dm2}
\begin{eqnarray}
\epsilon_{ij}\left(\omega,\mathbf{k}\right) =\delta _{ij}-\frac{4\pi
e^{2}}{m\omega^{2}}\int
\frac{d^{3}\mathbf{p}}{\gamma}\frac{\gamma^{2}\left(\omega-\mathbf{k\cdot
v}\right)^{2}}{\gamma^{2}\left(\omega-\mathbf{k\cdot v}\right)
^{2}-Q_{r}^{2}}\nonumber\\
\times\left[\delta
_{ij}+\frac{k_{i}v_{j}+k_{j}v_{i}}{\left(\omega-\mathbf{k\cdot
v}\right)}+\frac{\left(k^{2}-\omega^{2}/c^{2}\right)
v_{i}v_{j}}{\left(\omega-\mathbf{k\cdot v}\right)^{2}}\right]
f\left(\mathbf{p}\right), \label{35}
\end{eqnarray}
where $\mathbf{p}=\gamma m\mathbf{v}$, $\gamma=\left(
1-v^{2}/c^{2}\right)^{-1/2},$ and $f\left(\mathbf{p}\right)
=2\bar{n}\left(\mathbf{p}\right)/\left( 2\pi \hbar \right)^{3}$ with
$\bar{n}\left(\mathbf{p}\right) $ being the sum of the occupation
numbers for electrons and positrons and
\begin{equation}
Q_{r}=\frac {\hbar}{2m} \left(\frac {\omega^{2}}{c^{2}}-k^{2}\right).
\label{36}
\end{equation}

It is instructive to consider the first denominator of the integrand in Eq.~(\ref{35}) the zeroes
of which contain the resonance condition for the interaction of electrons (positrons) with the electromagnetic wave:
\begin{equation}
\gamma^{2}\left(\omega-\mathbf{k\cdot v}\right)^{2}-Q_{r}^{2} =
[\gamma\left(\omega-\mathbf{k\cdot v}\right)+Q_{r}][\gamma\left(\omega-\mathbf{k\cdot v}\right)-Q_{r}].
\label{eq:rel-resonance}
\end{equation}
Even though the electromagnetic field is treated classically the
zeroes of the two factors can be understood as arising from the
emission and absorption of a field quantum by the particles. Thereby
the particle energy and momentum change from $E$ to $\acute{E}$ and
$\mathbf{p}$ to $\mathbf{\acute{p}}$ by the discrete amount of
$\hbar \omega $ and $\hbar \mathbf{k}$, respectively, i.e.
$\acute{E} =E \mp \hbar \omega $ and
$\mathbf{\acute{p}}=\mathbf{p}\mp\hbar \mathbf{k}$, where
$E^{2}=\mathbf{p}^{2}c^{2}+m^{2}c^{4}$ and
${\acute{E}}^{2}={\mathbf{\acute{p}}^{2}} c^{2}+m^{2}c^{4}$. For
non-relativistic particle velocities $(\gamma\approx1)$, the
resonance condition (\ref{eq:rel-resonance}) becomes
\begin{equation}
\omega-\mathbf{k\cdot v} \mp \frac {\hbar}{2m} \left(\frac
{\omega^{2}}{c^{2}}-k^{2}\right) =0.  \label{36a}
\end{equation}
There other interesting limit is the classical limit. Then the
resonance condition (Cherenkov condition) is simply
$\omega-\mathbf{k\cdot v} = 0$. Clearly, this limit is recovered by
putting $Q_r \to 0$ which amounts to neglecting quantum effects
(terms proportional to $\hbar$). The quantum correction $Q_r$ to the
classical case is frequently called ``quantum recoil'' \cite{dm2}
although this is slightly misleading\footnote{The energy and
momentum balance that includes the absorption and emission of
photons has been written above and does not contain any additional
``recoil'' contribution.} \index{quantum recoil}

It is important to note that in a strictly non-relativistic
treatment, where one uses the dispersions $E=\mathbf{p}^{2}/2m$, the
term $\omega ^{2}/c^{2}$ doesn't appear in the expressions
(\ref{36}) and (\ref{36a}), showing that a non-relativistic
treatment is valid only for $\omega^{2}\ll \mathbf{k}^{2}c^{2}$ -- a
fact that is well-known in the theory of plasma oscillations but
still often ignored, see e.g. Ref.~\cite{dm2}.
From Eq. (\ref{35}), the dispersion relation for longitudinal
electron waves can be found as before from vanishing of the longitudinal part
of the tensor, $\varepsilon _{ij}(\omega ,\mathbf{k})=0$.
\\
{\bf Damping of plasma waves.} \index{electron plasma wave} The use
of QHD neglects certain kinetic effects such as Landau
damping.\index{wave damping} This effect is easily treated taking
into account that the dielectric function, Eq. (\ref{30}), is
complex since it includes a small imaginary correction to the
frequency (see footnote above). So far we did only consider its real
part.

In general, the poles at $v=\omega/k\pm \hbar k/2m$ have both a real and an
imaginary part, so the integration has to be performed using the Landau
pole integration in the complex velocity plain (analytic continuation is assumed \cite{mb6}),
\begin{equation}
\epsilon\left(\omega,k\right) =1-\frac{m\omega _{p}^{2}}{n_{0}\hbar
k^{2}}\left[\int_{C_{+}}\frac{f_{0}\left( v\right) }{k\left( v-\hbar
k/2m\right)-\omega}dv-\int_{C_{-}}\frac{f_{0}\left(v\right)
}{k\left(v+\hbar k/2m\right)-\omega}dv\right]=0,  \label{37}
\end{equation}
where the integration is performed with Landau contours $C_{\pm }$
passing under the poles at $v=\omega /k\pm \hbar k/2m$. Equation
(\ref{37}) is a useful starting point for the discussion of the
quantum Landau damping just like the collisionless damping in
classical plasmas \cite{yk2,fh1}. Adopting the procedure similar to
the classical plasmas, and assuming small damping (or growth)
rate $|\omega _{i}|\ll \omega$, we obtain \cite{mb2,mb6}
\begin{equation}
\omega_{i}=\frac{\pi\omega_{p}^{3}}{4n_{0}k^{2}}\left[
\frac{f_{0}\left(\omega/k+\hbar k/2m\right)-f_{0}\left(\omega
/k-\hbar k/2m\right)}{\hbar k/2m}\right]. \label{38}
\end{equation}
In the limit $\hbar\rightarrow 0,$ the known classical relation is recovered,
\begin{equation}
\omega_{i}=\frac{\pi\omega
_{p}^{3}}{2n_{0}k^{2}}\frac{df_{0}}{dv}\mid_{v=\omega/k},\label{39}
\end{equation}
which shows that (\ref{38}) can be considered as a finite-difference
generalization of the classical expression (\ref{39}).

Generalizing (\ref{37}) to three dimensions, the dispersion equation
becomes
\begin{equation}
\epsilon=1-\frac{m\omega_{p}^{2}}{n_{0}\hbar k^{2}}\left[
\int_{C_{+}}\frac{f_{0}\left(\mathbf{v}\right)}{k\left( v_{z}-\hbar
k/2m\right)-\omega }d\mathbf{v}-\int_{C_{-}}\frac{f_{0}\left(
\mathbf{v}\right)}{k\left( v_{z}+\hbar k/2m\right)-\omega
}d\mathbf{v}\right]=0, \label{40}
\end{equation}
where $f_{0}\left(\mathbf{v}\right) $ is the equilibrium Wigner
distribution function, the Landau contours are passing below the
poles lying at $v_{z}=\omega /k\pm \hbar k/(2m)$, and the coordinate
system is chosen such that the wave vector points in $z$-direction,
$\mathbf{k}=(0,0,k)$. Introducing $f_{0z}\left( v_{z}\right) =\int
dv_{x}dv_{y}f_{0}\left( \mathbf{v}\right),$ Eq.~(\ref{40}) can be
integrated over the perpendicular velocity components, leading to
\begin{equation}
\epsilon=1-\frac{m\omega_{p}^{2}}{n_{0}\hbar k^{2}}\left[
\int_{C_{+}}\frac{f_{0}\left(v_{z}\right) }{k\left(v_{z}-\hbar
k/2m\right)-\omega}dv_{z}-\int_{C_{-}}\frac{f_{0}\left(
v_{z}\right)}{k\left( v_{z}+\hbar k/2m\right)-\omega}dv_{z}\right]
=0. \label{41}
\end{equation}
This result is formally the same as (\ref{37}), which allows one to write
for the damping/growth rate in the classical limit, analogous to (\ref{39}),
\begin{equation}
\omega_{i}=\frac{\pi\omega_{p}^{3}}{2n_{0}k^{2}}
\frac{df_{0z}}{dv_{z}}\mid_{v_{z}=\omega
/k}. \label{42}
\end{equation}
Damping or growth depend on the sign of the derivative of the
projected equilibrium Wigner function. In equilibrium, the
distribution function is monotonically decaying with momentum and
$\omega_i$ is negative, corresponding to damping of the wave. In
non-equilibrium, the situation can be opposite.\footnote{Note that
the existence of instabilities depends on the system dimensionality.
While a monotonic increase of $f_0$ leads to an instability in a
one-dimensional and two-dimensional system, this is not the case in
a spherically symmetric 3D system \cite{bonitz_pop94}.}

To apply the above result to a degenerate plasma, it is useful to
start from the finite-temperature case because the zero-temperature
distribution has a singular derivative. Therefore, consider the
Thomas-Fermi distribution \cite{wf}
\begin{equation}
f_{0}\left( \mathbf{v}\right)
=\frac{\alpha}{\exp\left[\beta\left(\frac{mv^{2}}{2}-\mu \right)
+1\right]}, \label{43}
\end{equation}
where $v^{2}=$ $v_{x}^{2}+v_{y}^{2}+v_{z}^{2},$  $\beta=\left(
k_{B}T\right)^{-1},$ and the normalization constant $\alpha=2\left(
m/2\pi\hbar\right)^{3}.$ When the temperature $T$ approaches zero,
$\mu$ approaches to the Fermi energy $E _{F}=mv_{F}^{2}/2.$ Then the
integration over the perpendicular velocity components leads to
\begin{equation}
f_{0z}\left( v_{z}\right)=\frac{2\pi\alpha }{m\beta}\ln\left[
1+\exp\left[\beta \left(\mu-\frac{mv_{z}^{2}}{2}\right)\right]
\right], \label{44}
\end{equation}
which has a bell shaped profile. Then, by employing (\ref{42}) for
the damping, we obtain
\begin{equation}
\omega_{i}=\frac{\pi^{2} \alpha \omega_{p}^{4}}{n_{0}k^{3}}\left[
1+\exp\left[\beta \left(\frac{m\omega_{p}^{2}}{2k^{2}}-\mu
\right)\right]\right]^{-1}, \label{45}
\end{equation}
where in deriving (\ref{45}), the replacement $\omega \simeq
\omega_{p}$ was done. Upon analyzing (\ref{45}), we can see that for
very low temperature (large $\beta$), $\mu \simeq E_{F}$. Then for
$\omega_{p}/k>v_{F},$ there will be no damping because the wave
phase velocity lies in a region where there are no particles.
Therefore, high-frequency electron plasma oscillations of a
degenerate plasma at very low temperature remain undamped (in the
absence of particle collisions \cite{aa}). On the other hand, when
$\omega_{p}/k<v_{F},$ the exponential term in (\ref {45}) becomes
zero for very large $\beta $ and damping is significant\footnote{The
region where the imaginary part of the dielectric function is
non-zero and damping occurs at $T=0$ is called ``pair continuum''
since in this region the plasma wave loses energy by processes where
electron-hole pairs are created even when no collisions are taken
into account, see e.g. \cite{mb6}. At finite temperature, there
always exist particles with high velocity, so the damping is always
non-zero.} which cannot be taken into account in the QHD
description. So the long wavelength assumption $\left(\omega_{p}/k >
v_{F}\right)$ must hold in the QHD application to degenerate plasmas
to avoid damping of waves.
\subsubsection{Streaming instabilities}\label{ss:streaming}

Considering the one-stream plasma case with a single pure quantum
state\footnote{Recall that the Pauli principle prohibits that
several electrons move with exactly the same velocity. In reality
even in an electron beam the particles have a finite velocity spread
$\Delta v$, and the present model is to be understood as the limit
of small velocity spread, $\Delta v/u_0 \ll 1$.} with equilibrium
solutions $n=n_{0}$ and $u=u_{0}$ at $\varphi=0$, the Fourier
decomposition of the perturbed quantities in (\ref{13}), (\ref{16})
and (\ref{24}) leads to the dielectric function
\begin{equation}
\epsilon\left(\omega,k\right) =1-\frac{\omega_{p}^{2}}{\left(\omega
-ku_{0}\right)^{2}-\hbar^{2}k^{4}/4m^{2}}, \label{46}
\end{equation}
where the term $ku_{0}$ just represents a Doppler shift and
$v_{F}\ll u_{0} $ is assumed. Charge neutrality is provided by the
motionless background ions. Here, the frequency $\omega$ is always
real, and the oscillations are stable and undamped \cite{fh2}. If
the effect of quantum statistics is included in the momentum
equation, the dielectric function changes to
\begin{equation}
\epsilon\left(\omega,k\right) =1-\frac{\omega_{p}^{2}}{\left(
\omega-ku_{0}\right)^{2}-k^{2}v_{F}^{2}-\hbar ^{2}k^{4}/4m^{2}}.
\label{47}
\end{equation}
When two counter-streaming beams\index{two-stream instability} of
electrons are considered at equilibrium with streaming velocities
$\pm u_{0}$ such that $u_{1}=-u_{2}=u_{0}$, $n_{1}=n_{2}=n_{0}/2,$
and $\varphi =0,$ the dielectric function\index{dielectric function}
becomes
\begin{equation}
\epsilon\left(\omega,k\right)=1-\sum_{a=\pm 1}\frac{\omega
_{p}^{2}/2}{\left(\omega - a ku_{0}\right)^{2}-k^{2}v_{F}^{2}-\hbar
^{2}k^{4}/4m^{2}}, \label{48}
\end{equation}
which shows a Doppler shifted spectrum \cite{gm2} where the
dispersion relation\index{dispersion relation} is obtained from
$\epsilon\left(\omega,k\right)=0.$ When the solution for
$\omega^{2}$ is obtained, two branches are found, one of which is
always positive giving stable oscillations. The other solution is
negative $\left(\omega^{2}<0\right)$ which shows
\begin{equation}
\left[F^{2} K^{2}-4\left(1-u_{F}^{2}\right)\right] \left[F^{2}
K^{4}-4\left(1-u_{F}^{2}\right)K^{2}+4\right] <0,\label{49}
\end{equation}
\begin{figure}
\centering
\includegraphics [scale=1]{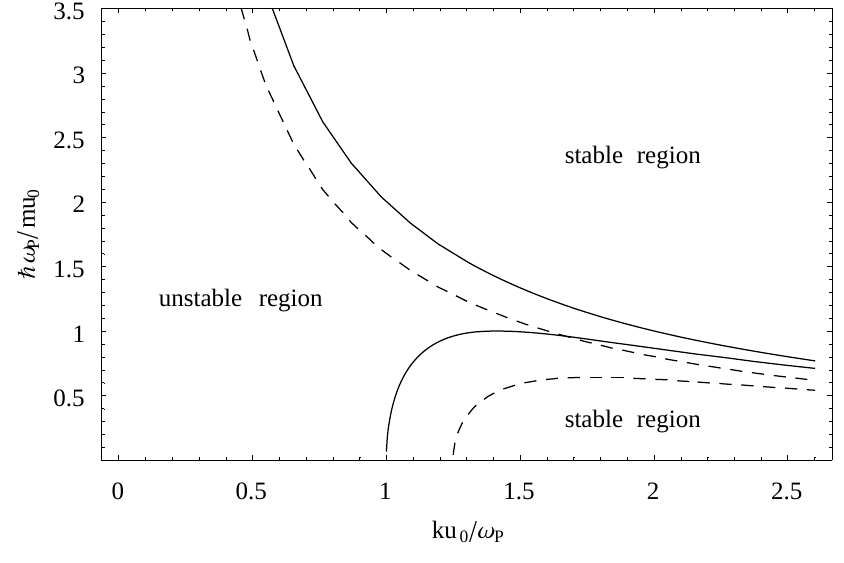}
\caption{Two-stream instability analysis in a quantum plasma by
using normalized parameters. The area enclosed by the solid curves
along the vertical axis shows an unstable region when $u_{F}=0$.
With increase in $u_{F}$, the region shows a shifting as seen by the
dashed lines corresponding to $u_{F}=0.6.$}\label{fig:4}
\end{figure}
where the rescaled variables are $F =\hbar \omega_{p}/mu_{0}^{2}$,
$K =ku_{0}/\omega _{p}$, and $u_{F}=v_{F}/u_{0}$. For $u_{F}<1$, a
bifurcation in (\ref{49}) is seen for $F =1-u_{F}^{2}.$ If
$F\geq1-u_{F}^{2}$, the second factor of the inequality is always
positive for $F ^{2}K^{2}<4\left( 1-u_{F}^{2}\right)$ which gives
rise to an instability. Similarly, if $F <1-u_{F}^{2},$ instability
occurs if either $0<F ^{2} K^{2} <2\left(1-u_{F}^{2}\right)
-2\sqrt{\left(1-u_{F}^{2}\right)-F ^{2}}$, or
$2\left(1-u_{F}^{2}\right) -2\sqrt{\left( 1-u_{F}^{2}\right)^{2}-F
^{2}}<$ $F^{2} K^{2}<2\left( 1-u_{F}^{2}\right)$. The limit
$\hbar\rightarrow 0$ leads to $K^{2}<1$ which is the classical
instability criterion. The stability/instability region can be seen
in Fig. \ref{fig:4} with a shift in the presence of nonzero Fermi
velocity.

The one and two stream cases show the main features of the
oscillation spectrum. However, when generalized to a larger number
of streams, the velocity spread, the coherence and resonant
contribution as well as collision between the groups of particles
lead to additional damping or dephasing, and a kinetic treatment is
required.

\subsubsection{Longitudinal ion waves}\index{ion wave}

When dealing with electrostatic oscillations having a frequency
close to the electron plasma frequency, the response of the ion
motion is very weak and does not need to be taken into account.

That's why, the ions were considered motionless in the previous section, forming a
neutralizing background. However, when the wave frequency is less
than the ion plasma frequency, $\Omega_{p}=(m/M)^{1/2}\omega_{p}$,
the dynamics of both species have to be taken into account
[$M$ is the ionic mass]. In a completely degenerate
two-component electron-ion quantum plasma, the Fermi energy of the
lighter species (electron) is larger than that of the ions due to smaller
electron mass, $E_F\sim m^{-1}$. Similarly, the de Broglie wavelength
scales as $m^{-1/2}$ and the degeneracy parameter as $\chi \sim m^{-3/2}$,
thus the ion degeneracy is much smaller than the one of the electrons. So, we
will continue to consider the ions classical.

In the case of classical plasmas, the longitudinal ion oscillations give rise to
the low-frequency ion-acoustic wave which is modified in quantum plasmas,
and a quantum ion-acoustic mode appears \cite{fh3}. The wave
dispersion relation of longitudinal ion waves in homogenous
electron-ion plasmas is given by the zeroes of the longitudinal
dielectric function,
\begin{equation}
1+\chi_{e}^{lo}\left(\mathbf{k},\omega\right)+\chi
_{i}^{lo}\left(\mathbf{k},\omega \right) =0,  \label{50}
\end{equation}
where $\chi_{e}^{lo}$ and $\chi_{i}^{lo}$ are the electron and ion
susceptibilities, respectively. For low phase velocity, it follows $\omega\ll
kv_{Fe},$  $\chi_{i}^{lo}\left(\mathbf{k},\omega \right) =-\Omega
_{p}^{2}/\omega^{2}$, resulting in
\begin{equation}
1+\frac{\omega_{p}^{2}}{k^{2}v_{F}^{2}+\hbar
^{2}k^{4}/4m^{2}}-\frac{\Omega_{p}^{2}}{\omega^{2}}=0. \label{51}
\end{equation}
The result for the dispersion can be written as
\begin{equation}
\omega^{2}=\frac{\Omega_{p}^{2}}{1+\aleph}, \label{52}
\end{equation}
where $\aleph =\omega_{p}^{2}/\left(k^{2}v_{F}^{2}-\hbar
^{2}k^{4}/4m^{2}\right).$ For $\aleph\gg 1,$ (\ref{52}) reduces to
\begin{equation}
\omega\simeq kc_{q}\left(1+\frac{\hbar^{2}k^{4}}{4m^{2}\omega
_{p}^{2}}\right)^{1/2}, \label{53}
\end{equation}
where $c_{q}=\left(E_{F}/M\right)^{1/2}$ is the speed of linear
electrostatic ion waves in a quantum plasma (the so-called quantum
ion-acoustic wave). Note that the relation (\ref{53}) is for a
non-relativistic ideal plasma at $T=0.$ In the classical limit,
$\hbar \rightarrow 0,$ $E_{F}\ll E_{T},$ Eq. (\ref{53}) corresponds
to the dispersion relation of the usual ion-acoustic wave in a
thermal plasma.

When using the QHD equations, the momentum equation (\ref{15}) for
ions can be written as
\begin{equation}
\left(\frac{\partial}{\partial t}+u_{i}\frac{\mathbf{\partial
}}{\partial x}\right) u_{i}=\frac{q_{i}E}{M}, \label{54}
\end{equation}
with $u_{i}$, and $q_{i}$ being the ion velocity, and ion charge,
respectively. The last two terms in (\ref{15}) are ignored for ions
due to smallness of ionic quantum effects. Similarly, the electron
inertia can be neglected in the limit $m/M\ll 1$. The space charge
electric field $E$ $=-\partial \varphi /\partial x$ couples ions
with the electrons.

If there exists a drift between electrons and ions in a quantum
plasma, the Buneman mode\index{Buneman instability} appears
\cite{fh4} just like the classical plasmas \cite{buneman}. Taking
into account collision effects in the electron and ion momentum
equations with $\nu_{en}\left(\nu_{in} \right)$ being the collision
frequencies of an electron (ion) with neutrals\footnote{This
approximation assumes a weakly ionized plasma where neutrals
dominate and, hence, collisions with neutrals play the main role.
This is the case at low temperature and not too high densities below
the Mott point.}, the dispersion relation in the reference frame of
the drifting ions becomes
\begin{equation}
1-\frac{\omega_{p}^{2}}{\left(\omega -kv_{0}\right)\left(\omega
-kv_{0}+i\nu_{en}\right)-\hbar^{2}k^{4}/4m^{2}}-\frac{\Omega
_{p}^{2}}{\omega\left(\omega+i\nu_{in}\right)}=0, \label{55}
\end{equation}
where the relative electron-ion equilibrium drift velocity in the
presence of a static electric field $E_{0}$ is
\begin{equation}
v_{0}=-eE_{0}\left( \frac{1}{m\nu _{en}}+\frac{1}{M\nu _{in}}\right)
. \label{56}
\end{equation}
For very low frequencies, $\omega \ll \nu _{in}\ll kv_{0}$, $\nu
_{en}\ll kv_{0}$, the dispersion equation predicts that the mode is
unstable under the condition
\begin{equation}
\omega _{p}^{2}>k^{2}v_{0}^{2}-\frac{\hbar^{2}k^{4}}{4m^{2}}>0,
\label{57}
\end{equation}
because $\Im (\omega)>0 $, otherwise it is damped. The small
wavelength oscillations are stable due to the presence of quantum
effects. For a slow temporal dynamics, appropriate rescaling of the
parameters gives rise to electron momentum equation of the form
\begin{equation}
v_{e}\frac{\partial v_{e}}{\partial x}
=-E+\frac{F^{2}}{2}\frac{\partial }{\partial
x}\left(\frac{\partial^{2}\sqrt{n_{e}}/\partial
x^{2}}{\sqrt{n_{e}}}\right), \label{57e}
\end{equation}
provided $\nu_{en} \ll \omega _{p}$ with $ F=\hbar\omega_{p}/m
v_{0}^{2}$ being the dimensionless parameter which measures the
contribution of quantum potential.\index{Bohm potential} The
linearization of the normalized set of equations around homogenous
equilibrium leads to the dispersion relation with imaginary part of
the frequency
\begin{equation}
\omega _{i}=\frac{k^{2}\left( 1-F ^{2}k^{2}/4\right) }{1-k^{2}+F
^{2}k^{4}/4}, \label{58}
\end{equation}
\begin{figure}
\centering
\includegraphics [width=8cm]{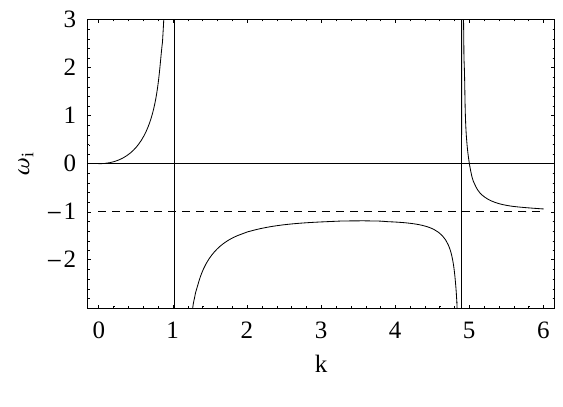}
\caption{Instability growth rate from Eq. (\ref{58}) is plotted as a
function of normalized wave number $ k $ for $F=0.4$. Two asymptotic
values of $k$ are denoted by $k_{1}$ and $k_{2}.$ Instability occurs
at $ 0 < k < k_{1} $ and $ k_{2} < k < k_{3} $. The growth rate
$\omega_{i}\rightarrow -1$ as $k\rightarrow \infty.$ in both the
cases of zero and nonzero $F$.} \label{fig:5}
\end{figure}
{
where $\Omega _{p} \ll \nu _{in}$. When the quantum parameter $F =0$
(classical limit), $ \omega _{i}=k^{2}/\left( 1-k^{2}\right) $.
Then, a singularity appears at $k=1$ and a linear instability exists
for $0<k<1$. For non-zero $F $, various instability conditions
emerge for $F <1$, $F=1$, and $F>1$. We consider one example of the
growth rate in the weak coupling regime, $ 0<F<1 $, as shown in Fig.
\ref{fig:5}, which has two asymptotic values, $k_{1}$ and $k_{2}$,
given by
\begin{eqnarray}
k_{1}^{2} = \frac{2}{F^{2}}[1-\sqrt{1-F^{2}}], \label{k1}
\\
k_{2}^{2}= \frac{2}{F^{2}}[1+\sqrt{1-F^{2}}]. \label{k2}
\end{eqnarray}
The growth rate is positive for $ 0 < k < k_{1}, $ or $ k_{2} <
k < k_{3}, $ where
\begin{equation}
k_{3}^{2} = k_{1}^{2}+k_{2}^{2}= \frac{4}{H^{2}}. \label{k3}
\end{equation}
This case of Buneman instability in collisional quantum plasma is
formally similar to the two-stream instability already discussed in
Sec. \ref{ss:streaming} above.

\subsection{Nonlinear waves in quantum plasmas}

In the preceding section, we have discussed linearized QHD results
following the standard procedure of linearization. When the
amplitude of a wave in plasma grows sufficiently large, the
nonlinearities in the QHD equations grow and cannot be neglected any
more. This makes the system more complicated and its analysis more
difficult. The nonlinearities in plasmas may enter through various
processes like advection, trapping of particles in the wave
potential, the nonlinear Lorentz force, ponderomotive force, etc.
Sometimes, the nonlinearities in plasma contribute to the
localization of waves giving rise to different types of interesting
coherent structures, for instance solitary waves, shocks, vortices,
and so on.

Due to highly nontrivial physics involved in the nonlinear regime of
quantum plasmas, only a limited analysis has been done in QHD so
far. As was shown above the \textit{nonlinear Schr\"odinger (NLS)
equation} (\ref{22}) is equivalent to the QHD in many respects. It
has many properties characteristic of nonlinear waves, especially
localized modes and solitons, and beam-driven waves and
instabilities. The NLS equation\index{Schr\"odinger
equation!nonlinear} and its variants describe nonlinear physical
systems appearing in a wide spectrum of problems in (quantum)
plasmas\footnote{always assuming that the applicability conditions
of QHD are fulfilled.} and other fields, for example, in fluids and
water waves, ultrafast transmission systems, condensed matter
systems, and so on. NLS contains an additional nonlinear term in the
Schr\"odinger equation responsible for the nonlinear effects. The
solution of NLS Eq. (\ref{22}) also facilitates the verification of
numerical solvers and aids in the stability analysis. Discrete
nonlinear Schr\"odinger (DNLS) equations are also important in
discrete lattice models in nonlinear optics, condensed matter and
trapped Bose-Einstein condensates where a numerical evaluation is
straightforward using e.g. the Crank-Nicolson method, e.g.
\cite{bauch_springer}.

The third and fourth terms on the right-hand side of Eq. (\ref{22})
represent the nonlinearities associated with the nonlinear coupling
between the electrostatic potential and the quantum statistical
pressure associated with Fermi-Dirac statistics. Linearizing the
NLS-Poisson system gives the frequency spectrum (\ref{26}) where $n=
\left\vert\Psi \right \vert^{2}$. In equilibrium, ${\tilde V}\sim
n^{2}=$ constant, otherwise it is repulsive because ${\tilde V}$ is
derived from $p^{c}\left(n\right)$ which is related to the
dispersion of velocities in a Fermi gas. The NLS equation admits
modulational wave solutions and a stability analysis can be
performed by standard procedures. Depending upon the type of
nonlinearity, it is also capable to provide valuable information of
the quasi-stationary structures and nonlinear interaction mechanisms
of waves at various length scales \cite{ps}.

The fluid modeling of the nonlinear long-wave-short-wave interaction
in plasmas is provided by the \textit{Zakharov
equations},\index{Zakharov equations} first derived by Zakharov
\cite{zakharov}, which get modified in quantum plasmas
\cite{lg,fh5}. The derivation of the \textit{quantum Zakharov
system} follows a two-time scales analysis of the QHD equations
which becomes possible due to the presence of fast (Langmuir-type)
and slow (ion-acoustic) oscillations. The limitations of the model
are similar to QHD and the allowed wavelengths are $\lambda \gg
\lambda_{TF},$ or, equivalently, $k\lambda_{TF}\ll 1$. All QHD
variables, i.e., the electron (ion) macroscopic density
$n_{e}\left(n_{i}\right)$, velocity $u_{e}\left(u_{i}\right)$, and
electric field $E$  are separated into fast (subscript f) and slow
(subscript s) oscillatory components
\begin{eqnarray}
n_{e}\left(x,t\right)=n_{0}+n_{es}\left(x,t\right)+n_{ef}\left(x,t\right),
\label{z1}
\\
n_{i}\left(x,t\right)=n_{0}+n_{is}\left(x,t\right), \label{z2}
\\
u_{e}\left(x,t\right)=u_{es}\left(x,t\right)+u_{ef}\left(x,t\right),
\label{z3}
\\
u_{i}\left(x,t\right)=u_{is}\left(x,t\right), \label{z4}
\\
E\left(x,t\right)=E_{s}\left(x,t\right)+E_{f}\left(x,t\right).
\label{z5}
\end{eqnarray}
The slowly varying quantities are considered not significantly
changing over a period of oscillation whereas the fast quantities
assume zero average. In addition, the quasi-neutrality condition, $
n_{e}\approx n_{i}, u_{e}\approx u_{i}$ is assumed, and the high
frequency ion terms are disregarded due to the smallness of $m/M$.
This analysis is in the spirit of a classical plasma with the
inclusion of the quantum (Bohm) potential for a zero-temperature
electron (Fermi) gas. In 1D, this gives the quantum corrected
Zakharov equations
\begin{eqnarray}
i\frac{\partial E}{\partial t}+\frac{\partial^{2}E}{\partial x^{2}}
-H^{2}\frac{\partial ^{4}E}{\partial x^{4}} &=& nE, \label{64}
\\
\frac{\partial^{2}n}{\partial t^{2}}-\frac{\partial^{2}n}{\partial
x^{2}}+H^{2}\frac{\partial^{4}n}{\partial x^{4}} &=&
\frac{\partial^{2}\left \vert E\right \vert^{2} }{\partial x^{2}},
\label{65}
\end{eqnarray}
where $E$ and $n$ are normalized quantities describing the slowly
varying envelope field and plasma density, respectively, and quantum
corrections are included via the non-dimensional quantum parameter
$H=\hbar \Omega _{p}/E _{Te}$ with $\Omega _{p}$ being the ion
plasma frequency and $ E_{Te}$ the electron thermal energy. The
system can admit periodic, chaotic or similar states, and describes
nonlinear dynamics and instabilities. The extension of
(\ref{64})-(\ref{65}) to three dimensions makes the inclusion of
electromagnetic effects possible \cite{fh5}.

The nonlinear effects causes the distortion of waves in plasma .
Then, wave steepening can occur until some dispersive or dissipative
process kicks in which broadens the profile, in turn balancing the
nonlinear steepening. Haas and co-workers \cite{fh3} have attempted
to include the quantum effects in nonlinear ion wave excitations in
the small and large amplitude regimes in the QHD framework. In the
small amplitude limit, the 1D QHD equations (\ref{13}),
(\ref{15})-(\ref{16}) and (\ref{54}) reduce through multiscale
expansion with appropriate rescaling of parameters to some form of
\textit{Korteweg-de Vries (KdV)} equation.\index{Korteweg-de Vries
equation} Assuming $m/M\ll 1,$ the electron momentum equation
(\ref{15}) with the boundary condition $n_{e}=1$, $\varphi =0$ at
infinity leads to
\begin{equation}
\varphi=-\frac{1}{2}+\frac{n_{e}^{2}}{2}-\frac{\Gamma_{q}}{2}\left[
\frac{\partial _{x}^{2}\sqrt{n_{e}}}{\sqrt{n_{e}}}\right],
\label{60}
\end{equation}
where the non-dimensional quantum parameter $\Gamma_{q} = (\hbar
\omega_{p}/E_{F})^2$. For $\Gamma_{q}=0$, the charge density is
directly related to the potential by an algebraic equation. We now
introduce the slowly varying stretched coordinates
\begin{equation}
\xi=\epsilon^{1/2}(x-t),\hspace{1.3cm}\tau=\epsilon^{3/2}t,
\label{61}
\end{equation}
where $\epsilon$ is a small parameter proportional to the amplitude
of the perturbation which provides the basis of the scaling, merely
a convention. Then, expanding the state variables into a series
in powers of $\epsilon$, the low orders of $\epsilon$
result in a KdV type equation with quantum corrections given by
\begin{equation}
\frac{\partial \Phi}{\partial\tau}+2\Phi \frac{\partial
\Phi}{\partial\xi}+\frac{1}{2}\left(1-\frac{\Gamma_{q}}{4}\right)
\frac{\partial^{3}\Phi}{\partial \xi^{3}}=0,  \label{62}
\end{equation}
which admits solitary ion wave and periodic solutions. The function
$\Phi\left(\xi ,\tau \right)$ arises from the zero order solutions
and the boundary conditions.

It is important to discuss the features of the dispersive term in
Eq. (\ref{62}). The equation is obtained by employing the reductive
perturbative technique and the term containing the quantum
diffraction (coupling) parameter $\Gamma_{q}$ appears from the
electron equation (\ref{60}). For $\Gamma_{q}=4$, the dispersive
term in Eq. (\ref{62}) disappears. In this case, the quantum
diffraction exactly matches the classical dispersion term in the KdV
equation. Then, no soliton solution exists and only free streaming
is possible like for a free ideal classical fluid which eventually
produces a shock wave. Recall that the parameter $\Gamma_{q}$ is
related to the coupling strength of the plasma. Since the
applicability of QHD (as discussed in Sec.~\ref{ss:limitations}
above) is limited to weak coupling, $\Gamma_{q} < 1 $ the value
$\Gamma_q=4$ is certainly out of the scope of QHD. Nevertheless such
values are often considered and we present one example below to
illustrate the mathematical consequences. For $\Gamma_{q}\neq 4,$
setting the wave-frame position variable $\eta =\xi-V_{0}\tau$ with
constant wave phase velocity $V_{0}$ leads to akin to energy first
integral for a particle of unit mass in a (pseudo) potential well
(Sagdeev-like potential) whose localized solution depends upon
$\Gamma_{q}$ and $V_{0}.$ A general profile of the potential
$V\left(\Phi\right)$, for different values of $\Gamma_{q}$ and
$V_{0}>0$, is shown in Fig. \ref{fig:6} which exhibits the localized
(soliton) structure.
\begin{figure}[h]
\centering
\includegraphics [scale=1]{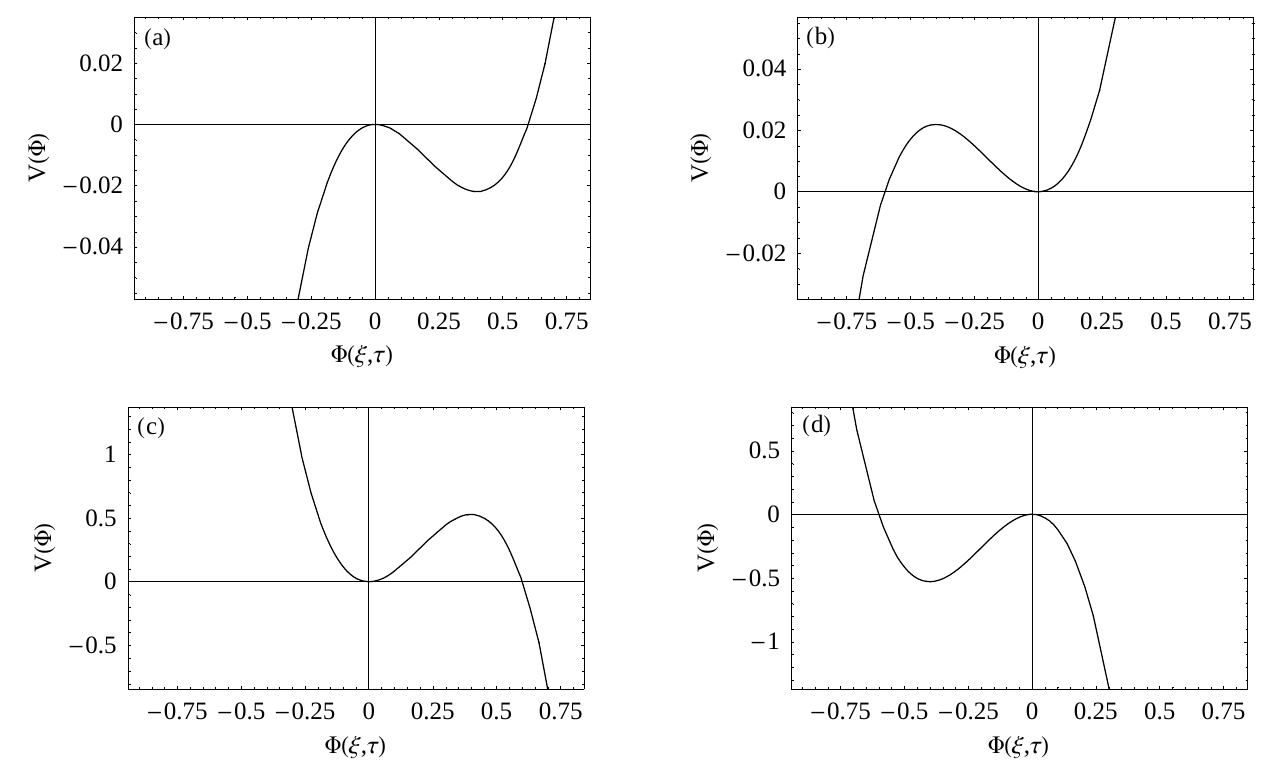}
\caption{The potential $V\left( \Phi \right) $ versus $\Phi \left(
\xi ,\tau \right) $ for arbitrary constant phase velocity $V_{0}$ is
shown for small and large values of $\Gamma_{q}.$ (a): $
\Gamma_{q}=0.1,$ $V_{0}=0.4,$ (b): $\Gamma_{q}=0.1,$ $ V_{0}=-0.4,$
(c): $\Gamma_{q}=4.1,$ $ V_{0}=0.4,$ (d): $ \Gamma_{q}=4.1,$ $
V_{0}=-0.4$. The conditions of solution; $ V^{^{\prime }}\left( \Phi
\right) =0$ at $\Phi =0$ and $\Phi =V_{0}$ are satisfied where prime
denotes the derivative with respect to $\eta. $} \label{fig:6}
\end{figure}
For $\Gamma_{q} < 4 $ and $V_{0}>0$, some algebra leads to a
solitary pulse solution of (\ref{62}) of the form
\begin{equation}
\Phi \left(\xi,\tau \right) =\frac{3V_{0}}{2}{\rm sech} ^{2}\left(
\frac{\eta }{\sqrt{\left( 4-\Gamma_{q}\right) /2V_{0}}}\right),
\label{63}
\end{equation}
as shown in Fig. \ref{fig:7} as a typical case. The pulse height
scales as $V_{0}$ whereas the pulse width as $V_{0}^{-1/2}$ which
also depends upon $\Gamma_{q}$. The larger amplitude pulses are
sharper and can propagate with higher speed.

\begin{figure}
\centering
\includegraphics [scale=1]{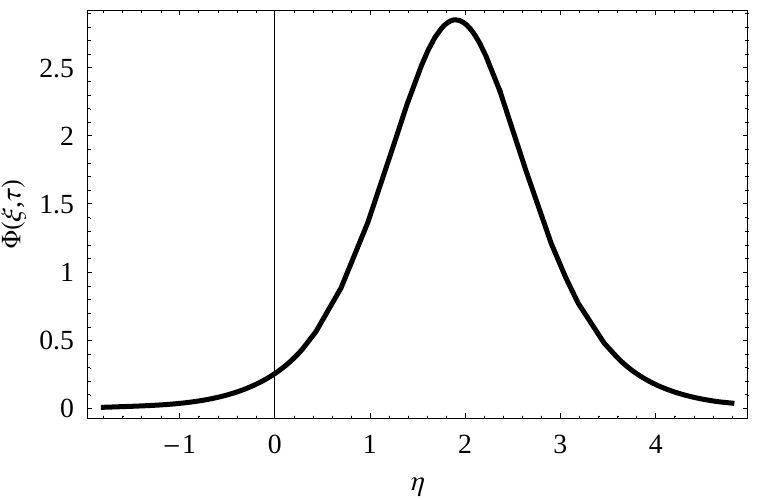}
\caption{The solitary pulse profile for $\Gamma_{q}=0.1$ and
arbitrary constant phase velocity $V_{0}>0.$ For such small values
of $\Gamma_{q}$, the effect on pulse width is negligibly small.}
\label{fig:7}
\end{figure}

For fully nonlinear large-amplitude localized ion waves, one
can define some non-dimensional parameter playing the role
of the Mach number. The wave form can be introduced via the
variable,
\begin{equation}
\xi=\left(x-Mt \right), \label{mach}
\end{equation}
with $M$ being the Mach number. Then, the QHD equations reduce to
the dynamical equations which can be written in the form of
conservation laws leading to solitary wave solutions depending upon
$\Gamma_{q}$ and $M$.

\subsection{Magnetized quantum plasmas}

So far, we have considered only the electrostatic case. The
inclusion of a magnetic field\index{quantum plasma!magnetized} leads
to a more general form of the QHD equations derivable from the
electromagnetic Wigner equation. Following the procedure similar to
the unmagnetized plasma case, the Madelung decomposition of the
ensemble wave functions allows to identify the classical and quantum
parts of the pressure dyad. Considering the statistical mixture of
$K-$ states $\psi _{i}=\psi _{i}\left( \mathbf{r} ,t\right)
,i=1,...,K$, such that the probabilities $p_{i}\geq0$, with
$\sum\limits_{i=1}^{K} p_{i}=1,$ each $\psi_{i}$ obey the
Schr\"odinger equation
\begin{equation}
i\hbar\frac{\partial\psi_{i}}{\partial t}=\frac{1}{2m}\left(
-i\hbar\mathbf{\nabla}-q\mathbf{A}\right)^2\psi_{i} + q\varphi \psi_{i},
\label{66}
\end{equation}
where the charge carriers have mass $m$ and charge $q$ under the
influence of self-consistent scalar and vector potentials $\varphi
\left( \mathbf{r},t\right) $ and $\mathbf{A}\left(
\mathbf{r},t\right)$, respectively, with choice of Coulomb gauge
$\mathbf{\nabla \cdot A}=0$. Then the one-particle Wigner
function\index{Wigner function} in terms of coordinate $\mathbf{r}$
and momentum $\mathbf{p}=m\mathbf{v}+q\mathbf{A}$ becomes
\begin{equation}
f_{W}\left( \mathbf{r},\mathbf{p},t\right)=\frac{1}{\left( 2\pi
\hbar \right)^{3}}\sum\limits_{i=1}^{K}p_{i}\int_{-\infty
}^{\infty}d\mathbf{s}\psi
_{i}^{\ast}\left(\mathbf{r}+\frac{\mathbf{s}}{2},t\right)e^{i\mathbf{p\cdot
s}/\hbar}\psi_{i}\left( \mathbf{r}-\frac{\mathbf{s}}{2},t\right),
\label{67}
\end{equation}
which leads to the evolution equation for $f_{W}$ -- the quantum
Vlasov equation -- obtained after a cumbersome calculation, for
details, see \cite{fh1}. Since, the complexity of the Wigner
function equation in the electrostatic case makes it very hard to be
fully examined, except for the simpler linear case. For a non-zero
magnetic field, the problem becomes even more challenging and the
analysis more difficult which motivates hydrodynamic description.

Introducing the moment equations in the usual way, the continuity
and the momentum transport equations in a magnetized plasma
following from the Wigner function equation can be written as
\begin{eqnarray}
\frac{\partial n}{\partial t}+\frac{\mathbf{\partial}\left(n\mathbf{u}\right)}{\partial x} &=& 0,
\label{68}\\
\frac{\partial \mathbf{u}}{\partial t}+\mathbf{u\cdot\nabla u}
&=& \frac{q}{m}\left(\mathbf{E}+\mathbf{u\times
B}\right)-\frac{1}{mn}\mathbf{\nabla}
p^{c}(n)+\frac{\hbar^{2}}{2m^{2}}\mathbf{\nabla}\left(
\frac{\nabla^{2}\sqrt{n}}{\sqrt{n}}\right),  \label{69}
\end{eqnarray}
where the closure assumption is made by defining a diagonal form of
the classical pressure dyad. Since the classical part of the pressure dyad
$p^{c}$ can be written as the sum of average velocity dispersions,
the diagonal isotropic form assumes the components
\begin{equation}
P_{ij}=\delta _{ij}p^{c}, \label{70}
\end{equation}
with $p^{c}=p^{c}\left( n\right)$ being a suitable equation of
state. For $\hbar \rightarrow 0,$ (\ref{69}) is just like the
momentum equation of classical fluids. Equations
(\ref{68})-(\ref{69}) together with Maxwell's equations constitute the
QHD model for magnetized plasmas where the limitations of the
electrostatic QHD equations are also valid in the present case. For
more subtle issues like gauge invariance of Wigner equation,
magnetohydrodynamic equilibria, the inclusion of spin
and strong field effects (Landau quantization) etc., see \cite{fh1}.
\begin{figure}[b]
\centering
\includegraphics [width=8cm]{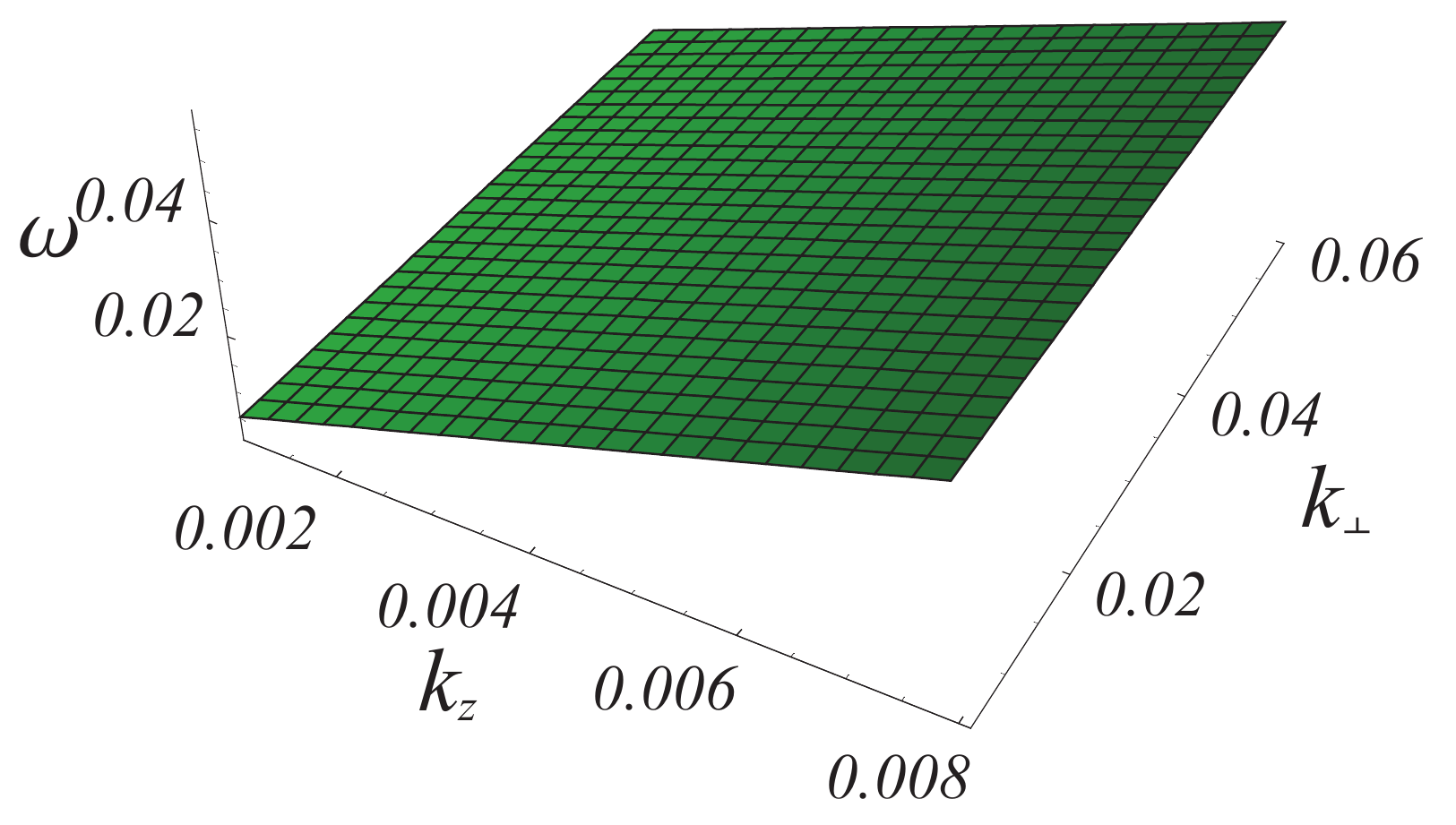}
\caption{Low-frequency electromagnetic mode in a strongly degenerate
electron plasma with immobile ions. The wave frequency is normalized
by $\omega_{c}$ and the wave numbers by $v_{F}/\omega_{c}.$}
\label{fig:8}
\end{figure}

\subsubsection{Electrostatic and electromagnetic low frequency modes}

We start from the QHD equations for a two-component dense uniform
magnetized plasma consisting of degenerate electrons and
non-degenerate ions. Assuming the dynamics of electrons with a
background of stationary ions embedded in a uniform magnetic field
$B_{0}\mathbf{\hat{z}}$, the low-frequency (in comparison with the
electron cyclotron frequency) electric and magnetic field
perturbations are defined as $\mathbf{E}=-\mathbf{\nabla
}\varphi-c^{-1}\left(\partial A_{z}/\partial
t\right)\mathbf{\hat{z}}$ and $\mathbf{B}_{\bot
}=\mathbf{\nabla}_{\bot }A_{z}\times\mathbf{\hat{z}},$ respectively.
From (\ref{69}), assuming low frequencies, $\omega \sim
|\partial/\partial t|\ll \omega_{c},ck$ $,$ but $\omega$ higher than
the ion plasma frequency $\Omega_{p}$, and ion cyclotron frequency,
$\Omega_{c}$, respectively, the linearized components of the
electron and ion fluid velocity parallel and perpendicular to
$\mathbf{\hat{z}}$ become
\begin{eqnarray}
\frac{\partial v_{jz}}{\partial t}=\delta _{j}\frac{e}{m}\left(
\frac{\partial \varphi _{j}}{\partial z}+\frac{1}{c}\frac{\partial
A_{z}}{\partial t}\right), \label{71}\\
\mathbf{v}_{j\bot}\simeq\frac{c}{B_{0}}\left( \mathbf{\hat{z}}\times
\mathbf{\nabla}_{\bot}\varphi_{j}+\delta_{j}\frac{1}{\omega_{c}}\frac{
\partial\mathbf{\nabla
}_{\bot}\varphi_{j}}{\partial t}\right), \label{72}
\end{eqnarray}
where $j=e (i)$ denotes electron (ion), $\delta_{e}=1$, $\delta
_{p}=-1$, and we denoted $\varphi_{j}=\phi
-\frac{\delta_{j}}{n_{0}e}\left(\frac{2E_{F}}{3}-\frac{\hbar
^{2}\nabla^{2}}{4m}\right) n_{j}$, with $n_{j}\ll n_{0}.$ The
dispersion relation for the shear electromagnetic
mode\index{electromagnetic mode} in this case is derived by
employing (\ref{71})-(\ref{72}), together with the linearized
continuity equation, Poisson's equation and Ampere's law which, upon
Fourier transformation, results in
\begin{equation}
\omega^{2}=\frac{c^{2}k_{z}^{2}k_{\bot}^{2}\left(1+\Lambda
_{q}^{2}k^{2}\right)}{\left(1+\lambda_{e}^{2}k_{\bot}^{2}\right)
\left[k^{2}+k_{\bot }^{2}\left( 1+\Lambda_{q}^{2}k^{2}\right)
\frac{\omega_{p}^{2}}{\omega_{c}^{2}}\right]}, \label{73}
\end{equation}
which shows the influence of the electron quantum statistical and
quantum diffraction effects [Fig. \ref{fig:8}], where $\lambda
_{e}=c/\omega _{p},$ $\Lambda _{q}=v_{q}/\omega _{p}$, and
$v_{q}=\sqrt{T_{q}/m}$ with
$T_{q}=\left(\hbar^{2}k^{2}/4m+2E_{F}/3\right)$ being the quantum
parameter in energy units playing the same role as effective
temperature in classical plasmas \cite{sk1}. The electromagnetic
mode (\ref{73}) ceases to exist for $k_{z}=0$.\\
Next, we consider the ion dynamics in the frequency regime $\omega
\ll \Omega_{c}.$ The ion perpendicular velocity component then
consists of the electric and ion polarization drifts, ${\bf v_{E}}$
and ${\bf v_{p}}$ respectively. Then the dispersion equation with
assumption of $k_{z}\ll k_{\bot}$ acquires the form
\begin{eqnarray}
\left[\left(1+K_{A}^{2}\right)\omega^{2}-\frac{c_{q}^{2}k_{z}^{2}}{\left(1+
\lambda_{e}^{2}k_{\bot}^{2}\right)}\left(1+K_{A}^{2}\right)\right]
\omega^{2}\nonumber\\
-\frac{\omega_{A}^{2}}{\left(1+\lambda_{e}^{2}k_{\bot}^{2}
\right)}\left[\left(1+\frac{c_{q}^{2}k^{2}}{\omega_{pi}^{2}}\right)\omega^{2}-c_{q}^{2}k_{z}^{2}
\right]=\frac{\rho_{q}^{2}k_{\bot}^{2}\omega_{A}^{2}}{\left(1+\lambda_{e}^{2}k_{\bot}^{2}\right)}\omega^{2},\label{74}
\end{eqnarray}\\
where $K_{A}=\frac{v_{A}}{ck_{\perp}/k},$ $v_{A}=B_{0}/\sqrt{4\pi
n_{0}M}$ is the speed of the Alfv\'{e}n wave, $c_{q}=\sqrt{T_{q}/M}$
is the speed of the electrostatic ion wave, and
$\rho_{q}=c_{q}/\Omega_{c}.$ The co-existing electrostatic and
electromagnetic modes [Fig. \ref{fig:9}] are well separated,
however, the difference of frequencies become lesser and lesser as
the magnetic field is increased.
\begin{figure}
\centering
\includegraphics [width=11cm]{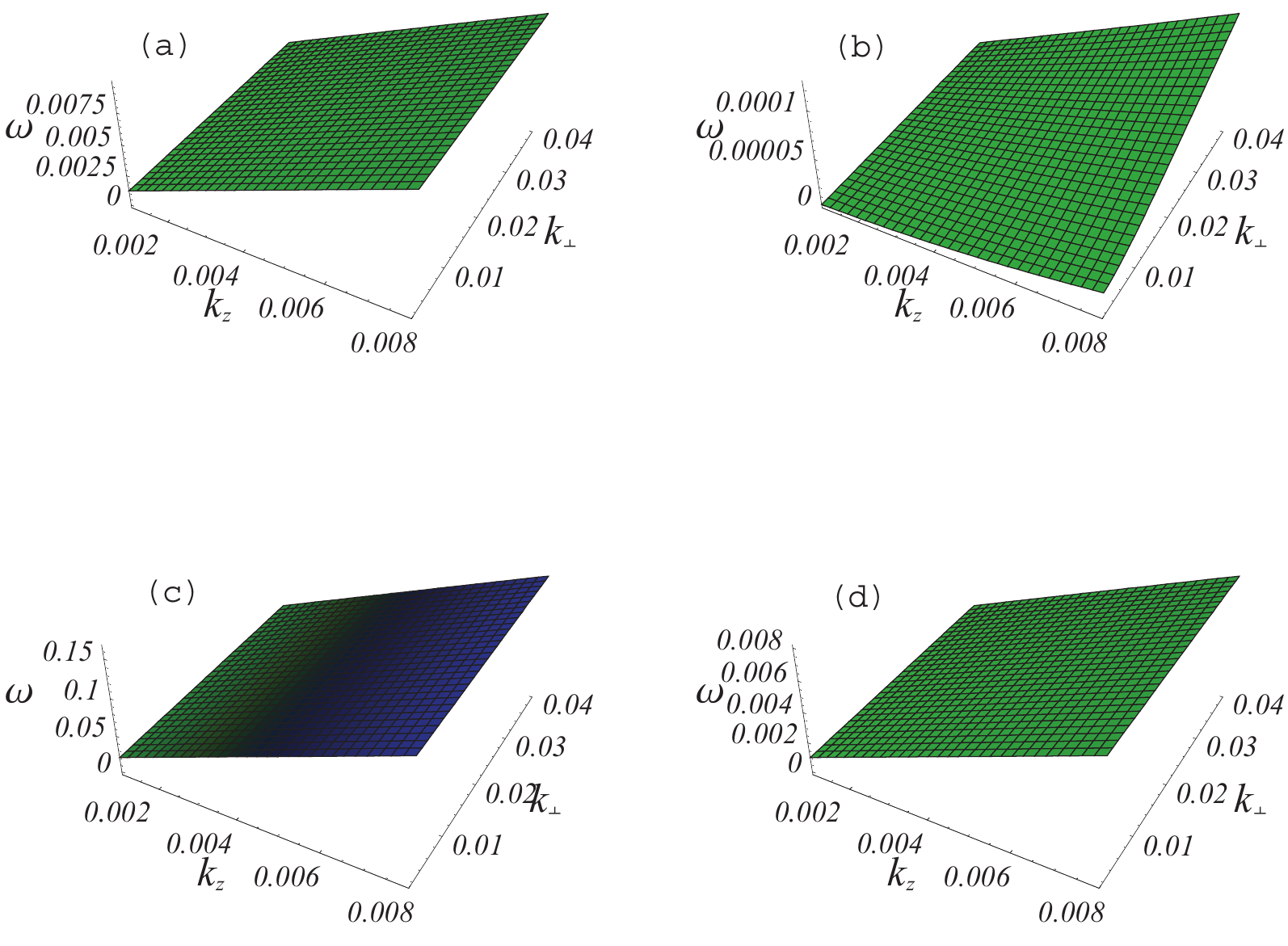}
\caption{Low-frequency linear electrostatic (a,c) and
electromagnetic (b,d) modes in a magnetized quantum plasma in the
absence (upper panel) and the presence (lower panel) of degeneracy
pressure for typical quantum plasma of white dwarf stars taken from
Table \ref {tab:2} with density $\sim 10^{26} cm^{-3}$ and magnetic
field $\sim 10^{8} G$. The dispersion relation shows that the
electrostatic (quantum ion-acoustic) mode can couple with the
electromagnetic (shear Alfv\'{e}n) mode in high magnetic field
regions where the difference of frequencies become small. The wave
frequency is normalized by $\Omega_{c}$ and wave numbers by
$v_{A}/\Omega_{c}.$ } \label{fig:9}
\end{figure}
In the limiting case, for $v_{A}\ll ck_{\bot }/k,$
$\lambda_{e}^{2}k_{\bot}^{2}\ll 1,$ (\ref{74}) reduces to
\begin{equation}
\omega^{2}=k_{z}^{2}v_{A}^{2}\left(
1+\rho_{q}^{2}k_{\bot}^{2}\right), \label{75}
\end{equation}
which shows the dispersive Alfv\'{e}n wave where the dispersion
comes from the electron quantum effects.

\subsubsection{Drift mode}
In the presence of gradients (inhomogeneities in density,
temperature, etc.), there may appear drift waves\index{drift wave}
in classical as well as quantum plasmas which play an important role
in transport of plasma particles and energy/momentum across the
magnetic field lines. Drift waves are low frequency waves in
comparison with the ion cyclotron frequency $\Omega _{c}$ with
perpendicular (with respect to the magnetic field) wave number
$k_{\perp}$ much larger than $k_{\parallel}$. For relatively large
$k_{\parallel}$, the drift wave can couple with the quantum ion
acoustic wave \cite{hs}. Consider a dense quantum plasma embedded in
a constant external magnetic field in z-direction possessing a
density inhomogeneity at equilibrium in the x-direction such that
$\mathbf{\nabla}n_{j0}=-(\frac{dn_{j0}}{dx}) \mathbf{\hat{x}}$, and
$\kappa_{jn}=|\frac{1}{n_{j0}}\frac{dn_{j0}}{dx}|$ = constant, with
$\kappa _{jn}\ll k_{\perp}$. Using the QHD equations with the
electric and magnetic field perturbations as given in the previous
section, the dispersion relation for drift waves in quantum plasma
becomes
\begin{eqnarray}
\left[(\omega-\omega_{q}^{\ast })\omega
-\frac{c_{q}^{2}k_{z}^{2}}{\left(1+\lambda_{e}^{2}k_{\bot
}^{2}\right)}\right]\omega^{2}\nonumber\\
-\frac{\omega_{A}^{2}}{\left( 1+\lambda_{e}^{2}k_{\bot
}^{2}\right)}\left[ \omega^{2}-\omega _{q}^{\ast}\omega
-c_{q}^{2}k_{z}^{2}\right]=\frac{\rho_{q}^{2}k_{\bot}^{2}\omega_{A}^{2}}{\left(1+\lambda_{e}^{2}k_{\bot}^{2}\right)}\omega^{2},
\label{76}
\end{eqnarray}
where $\omega_{A}=k_{z}v_{A}$ is the frequency of the Alfv\'{e}n wave,
$\omega_{q}^{\ast}=\mathbf{v}_{qD}.\mathbf{k}$ is the drift wave
frequency, $\mathbf{v}_{qD}=\frac{cT_{q}}{eB_{0}}\mathbf{\nabla}\ln
n_{0}\times \mathbf{\hat{z}}$ is the drift wave velocity, and
$T_{q}=\hbar^{2}k^{2}/4m$.\
\begin{figure}
\centering
\includegraphics [scale=1]{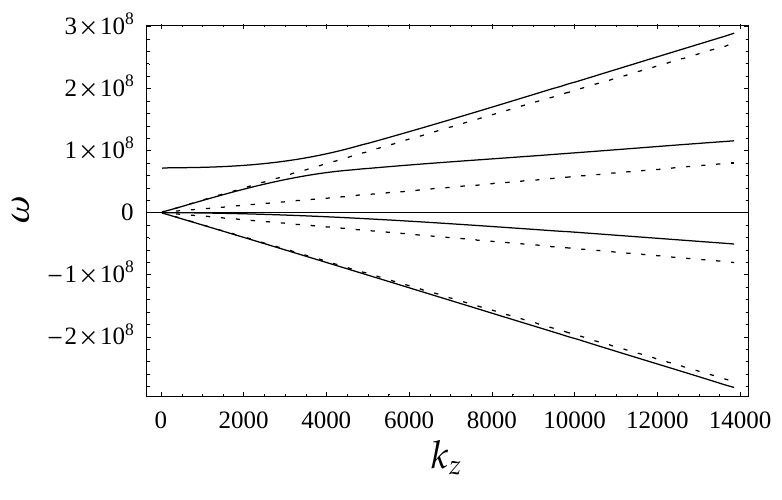}
\caption{The appearance of a drift mode in a magnetized non-uniform
quantum plasma due to density inhomogeneity. The frequencies of the
four low-frequency modes (in units of $s^{-1}$) are plotted vs. the
wave numbers $k_{z}$ for fixed $k_{y}$ (units of $cm^{-1}$) for a
dense astrophysical plasma with typical parameters selected from
Table \ref {tab:2} with density $\sim 10^{27} cm^{-3}$ and magnetic
field $\sim 10^{6} G$. One branch of the shear Alfv\'{e}n wave and
one branch of the electrostatic wave are influenced by the quantum
drift wave $\omega _{q}^{\ast }$ near $k_{z}\sim 4\times
10^{3}cm^{-1} $. The dashed lines represent $\pm k_{z}v_{A}$ (outer
lines) and $\pm k_{z}c_{q}$ (inner lines), respectively. From Ref.
\cite{hs}.} \label{fig:10}
\end{figure}
The above relation has an analogy with the classical drift wave
frequency which depends upon the equilibrium electron pressure
defined by the ideal gas law. However, both are very different
physically. The classical drift wave depends upon electron thermal
energy, but in a quantum plasma, the role of thermal energy is taken
over by the Fermi energy. In deriving (\ref{76}), the drift
approximation $|\partial_{t}|\ll \Omega _{c}$ is used in the limit
of small Fermi pressure. In Fig. \ref{fig:10}, the frequencies of
the four modes are plotted against $k_{z}$. One branch of the shear
Alfv\'{e}n wave and one branch of the electrostatic wave are
influenced by the quantum drift wave $\omega_{q}^{\ast }$ near
$k_{z}\sim 4\times 10^{3}cm^{-1}$. Since $T_{i}=0$ has been assumed,
therefore the second branch of the Alfv\'{e}n wave remains a
straight line in this figure. The second branch of the electrostatic
wave has also the effects of dispersion similar to the classical
case. The density matrix approach can also be used to study quantum
drift wave in two component inhomogeneous plasma in a strong
magnetic field under strong and weak quantum effects \cite{bs}.
Here, the problem is treated semiclassically with a modified Maxwell
distribution function in order to determine the quantum effects.
Such waves become unstable under some circumstances.\\

Concluding this section, let us briefly discuss for what systems the
present results can be relevant. In the QHD approach, the effect of
Fermi degeneracy and (quantum) Bohm potential\index{Bohm potential}
are the main quantum ingredients. The applications of nonlinear
waves with or without magnetic field can be found in dense
quasi-free electron gas and in the high density regimes relevant to
degenerate plasmas of dense astrophysical objects (regions of white
dwarfs and neutron stars). Such densities are also expected in the
lab in next generation laser-plasma experiments. At such high
densities, $r_s$ will be well below unity and the QHD model may be
even better applicable.

Let us consider a typical example from Table \ref {tab:2}, with
density $n_{0}= 4.0\times10^{26} cm^{-3}$, $\bar{r}=
8.3\times10^{-10}$cm and $\lambda_{TF}= 7.2\times10^{-8} cm$. The
wave number $k$ should be well below $\bar{r}^{-1}$ for both
electronic and ionic perturbations. For low frequency perturbations,
if $k\simeq 10^{6} cm^{-1},$ the ion wave speed is $c_{q}\simeq
3.8\times 10^{7}$ms$^{-1}$ and the frequency, from Eq. (\ref {51})
is $\simeq 6\times 10^{13}s^{-1}$.
If we consider the soliton solution, Eq.~(\ref {63}), for such a
plasma, the effect of the quantum parameter $\Gamma _{q}$ is
vanishingly small since $\Gamma _{q}\leq 1$ for applicability of the
QHD approximation. Due to the very high density, the inter-particle
distances are very small. In the weakly nonlinear limit, a soliton
with a typical speed $\sim 0.1c_{q}$ shows very small amplitude and
width parameters.
For a magnetized plasma with high ambient magnetic field $ \left(
B_{0}\simeq 1\times 10^{7}G\right) $ and low frequency electrostatic
and electromagnetic perturbations$,$ dispersion equation Eq. (\ref
{74}) with $k_{z}/k_{\perp }\simeq 0.002$ reveals that the frequency
of electrostatic mode is $1.4\times 10^{13}s^{-1}.$ Similarly, the
frequency of the shear Alfv\'{e}n mode is $3.1\times 10^{9}s^{-1}$,
with $v_{A}=2.2\times 10^{5}$ms$^{-1}$ being the speed of the wave.
It should be noted that the modes are well separated for low
magnetic field but their frequencies get closer and closer as the
magnetic field increases.


%
\section{Interaction and spin effects in Quantum Plasmas}\label{s:interactions}
As was discussed above, QHD assumes an (almost) ideal electron Fermi
gas. There have recently been attempts to include exchange and
correlation effects in order to extend the validity range of QHD
which we briefly discuss below.

The first attempt to include exchange and correlation effects in QHD
phenomenologically was presented by Manfredi and co-workers in Ref.
\cite{nc}.\index{quantum hydrodynamics!with exchange-correlation}
Inspired by the procedure used in density functional theory (DFT)
they used an additional exchange-correlation functional, $V_{xc}$
\begin{equation}
V_{xc} = 0.985(e^2/\epsilon)n^{1/3}
[1+(0.034/a_Bn^{1/3})\ln(1+18.37a_Bn^{1/3})],
\end{equation}
in the momentum equation that gives rise to an additional
force on the electrons.
The authors performed comparisons with DFT simulations for electrons in condensed matter and observed reasonable agreement.

\subsection{Prediction of attractive forces between protons in quantum plasmas}
Using the QHD with the above mentioned potential $V_{xc}$ Shukla and
Eliasson \cite{shukla_prl12} considered the problem of the effective
potential of a proton embedded in a dense quantum plasma. The QHD
equations together with the Poisson's equation for the electrostatic
potential are
\begin{eqnarray}
\frac{\partial n}{\partial t}+\nabla\cdot(n\vec{\mathrm{u}}) &=& 0,
\nonumber\\
m_*\left(\frac{\partial \vec{\mathrm{u}}}{\partial t} +
\vec{\mathrm{u}} \cdot \nabla \vec{\mathrm{u}} \right) &=&
e\nabla\phi-n^{-1}\nabla P + \nabla V_{xc} + \nabla V_B,
\label{eq:momentum}\\
\nabla^2\phi &=& \frac{4\pi e}{\epsilon}(n-n_0)-4\pi Q\delta(\vec{r}),
\end{eqnarray}
where the positive test charge $Q$ is located at ${\bf r}=0$.
Quantum effects are taken into account as usual via the Bohm
potential $ V_B=(\hbar^2/2m_*)(1/\sqrt{n})\nabla^2\sqrt{n}$. The
pressure of the ideal Fermi gas at zero temperature,
$P=(n_0m_*v_*^2/5)(n/n_0)^{5/3}$, is used and complemented by the
exchange-correlation potential $V_{xc}$. Here, the following
definitions are used: $\epsilon$ denotes the relative dielectric
permeability of the material, $v_* = \hbar(3\pi^2)^{1/3}/m_*r_0$ is
the electron Fermi speed, $r_0 =n_0^{-1/3}$ is the Wigner-Seitz
radius, and $m_*$ is the effective mass of electron\footnotetext{The effective mass takes into 
account medium effects for the case of electrons in condensated matter 
systems. Here we will focus on electrons in a hydrogen plasma where 
$m_*$ coincides with the free electron mass.}.

Shukla and Eliasson linearized these equations, writing $n = n_0 +
n_1$ and $|n_1| \ll n_0 $. Neglecting dynamic effects in the
dielectric function,
$\varepsilon(\vec{\mathrm{k}},\omega)=\varepsilon(\vec{\mathrm{k}},0)$,
the electrostatic potential of a proton is given by
\begin{equation}\label{eq:phi_p}
\phi(\vec{r}) = \frac{Q}{2\pi^2}\int\frac{\exp(i\vec{\mathrm{k}}\cdot\vec{\mathrm{r}})}{k^2\varepsilon(\vec{\mathrm{k}})}d^3k.
\end{equation}
From the linearized QHD equations they obtained for the inverse dielectric function
\begin{equation}
\frac{1}{\varepsilon(k)} = \frac{(k^2/k_s^2)+\alpha k^4/k_s^4}{1+(k^2/k_s^2)+\alpha k^4/k_s^4},
\end{equation}
with the definitions
\begin{eqnarray}
\alpha = \hbar^2\omega_{pe}^2/4m_*^2(v_*^2/3+v_{ex}^2)^2, \\[1ex]
k_s = \omega_{pe}/\sqrt{v_*^2/3+v_{ex}^2}, \quad b_* =1/\sqrt{4\alpha-1}, \\[1ex]
v_{ex} = ({0.328e^2/m_*\epsilon r_0})^{1/2} {[1+0.62/(1+18.36 a_B {n_0}^{1/3})]^{1/2}}, \\[1ex]
k_r = (k_s/\sqrt{4\alpha})(\sqrt{4\alpha}+1)^{1/2}, \quad k_i =
(k_s/\sqrt{4\alpha})(\sqrt{4\alpha}-1)^{1/2}.
\end{eqnarray}
\begin{figure}[h]
\includegraphics[scale=1]{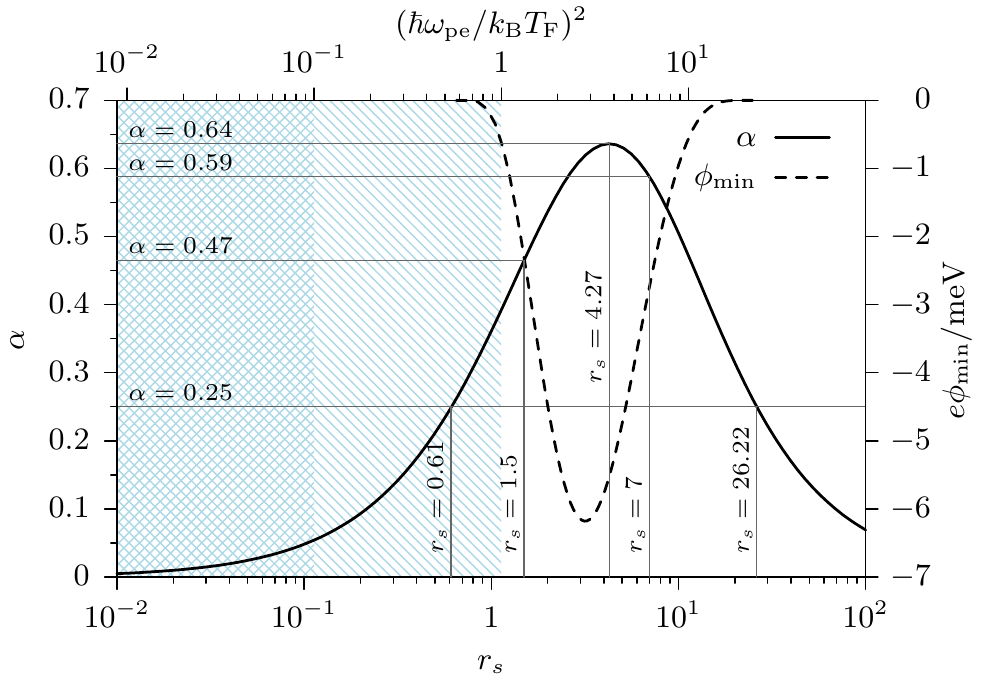}
\caption{ The Shukla-Eliasson parameter $\alpha$ (left axis) versus
quantum coupling parameters: the Brueckner parameter $r_s$ (bottom
axis) and $\Gamma_q$ (upper axis). The potential of a proton, Eq.\
(\ref{eq:phi_p}), derived by Shukla and Eliasson from linearized QHD
becomes attractive for $\alpha > 0.25$, corresponding to $26.22 \ge
r_s \ge 0.61$, at zero temperature. The two shaded areas denotes the
range of moderate (weak) coupling given by $(\hbar \omega_{\rm
pe}/k_{\rm B} T_{\rm F})^2 < 1$ ($<0.1$). Also shown is the depth of
the SE potential (right axis). From Ref. \cite{bonitz_phys-scr13}.
\index{Shukla-Eliasson potential}}
\label{fig:alpha}
\end{figure}
The parameter $\alpha$ is shown in Fig.~\ref{fig:alpha}. While for
$\alpha < 0.25$ the potential (\ref{eq:phi_p}) is always positive,
for $\alpha > 0.25$, it develops a negative (attractive) minimum. In
the latter case the potential is given by \cite{shukla_prl12}
\begin{equation}
\phi(\vec{r})=\frac{Q}{r}[\cos(k_i r)+b_*\sin(k_i r)]\, e^{-k_r r}.
\end{equation}
The maximum value of $\alpha$ is approximately $0.64$. Inserting all
parameters in the definition of $\alpha$, existence of a negative
potential in the linearized QHD is confined to a finite density
interval where $0.61 \le r_s \le 26.22$, see Fig.~\ref{fig:alpha}.
Here the standard quantum coupling (Brueckner) parameter has been
used, $r_s = {\bar r}/a_{\rm B}$, where ${\bar r}$ denotes the mean
inter-particle distance and $a_{\rm B}$ the Bohr radius. For weak
coupling, $r_s \ll 1$, linearized QHD does not predict a negative
potential.

In Ref. \cite{shukla_prl12}, based on the existence of a negative
minimum of the proton potential (\ref{eq:phi_p}), Shukla and
Eliasson claimed the discovery of a novel attractive force between
ions in dense quantum plasmas. They claimed that this potential
would lead to novel bound states and to a proton lattice. However,
as can be seen in Fig.~\ref{fig:alpha} where we also show the depth
of this potential, negative values occur only in the regime of
moderate coupling, i.e. way outside the validity range of QHD which
was discussed in Sec.~\ref{ss:limitations} above. We note that the
linearized version of QHD, obviously, is even less accurate. To
verify these strong claims, Ref. \cite{bonitz_comment} reported ab
initio density functional theory simulations. The DFT\index{density
functional theory} result for the effective potential of a proton in
dense hydrogen, indeed, was found to exhibit an attractive minimum,
in two cases: first, at low density, $r_s > 1.5$ there is a minimum
corresponding to binding of two hydrogen atoms into a molecule.
Second, at large distances, there are shallow oscillations of the
potential which are related to Friedel oscillations (originating
from the step character of the zero-temperature Fermi distribution).
No other cases of attractive potentials between protons were
observed in the simulations. Therefore, Ref.~\cite{bonitz_comment}
had to conclude that the predictions of Ref.~\cite{shukla_prl12} are
wrong.

The disagreement between linearized QHD and DFT was further
discussed in Refs. \cite{shukla_comment,bonitz_reply}. A careful
analysis of the applicability range of QHD and DFT shows that, from
its construction, DFT is always more accurate. The lesson to learn
from this is that the applicability limits of QHD should be taken
very seriously and clearly checked in any application.

\subsection{Spin effects in quantum plasmas}\label{ss:spin}
In recent years attempts have been made to extend the QHD to quantum
plasmas with spin effects.\index{spin effects}\index{quantum hydrodynamics} This is natural as
spin effects are always present for quantum particles. For the case
of plasmas with degenerate electrons, the effect of fermionic
statistics (spin $1/2$) has to be considered. The corresponding
extension of QHD to spin QHD (SQHD)\index{quantum hydrodynamics!with spin effects} can be found e.g. in
refs. \cite{marklund_prl07,mahajan_prl11,braun_prl12} and references
therein. These papers came to the conclusion that collective spin
effects can dominate the plasma dynamics which is derived from a
(possibly macroscopically large) spin magnetization current
\begin{equation}
 {\bf j}_{\rm spin} = \nabla \times 2n\mu_B {\bf S},
\label{eq:spin-current}
\end{equation}
where $n$ is the total electron density, $\mu_B$ the Bohr magneton and ${\bf S}$ the local average ``spin vector''.
For a high density, as is often the case in quantum plasmas and, assuming spin polarization (i.e. all spins
are aligned),
this current and the associated magnetization may become very large.

It has recently been pointed out \cite{spin_critique} that this prediction is in striking contrast
to standard condensed matter physics and experiments as well. In particular, the quantum theory
of magnetism, e.g. \cite{lali10,ashcroft-mermin,kittel} is well developed and does not predict
any such gigantic magnetizations. There it is known that the magnetization arises from unpaired
electron spins and is proportional to the density of spin up minus spin down electrons, $n_+ - n_-$.
However, due to Pauli blocking, at low temperature this difference vanishes, it is zero for an
ideal Fermi gas in the ground state. At finite temperature, a finite difference may exist which
scales as \cite{spin_critique}
\begin{equation}\label{eq:s+s-}
 \frac{n_+-n_-}{n} = O\left( \frac{T}{T_F} \right),
\end{equation}
i.e. in the magnetization not all electrons but only those in a
small layer (of order $T$) around the Fermi edge participate
resulting in the known moderate values for the spin magnetization of
real materials. Thus, at low temperatures when the system approaches
an ideal Fermi gas the spin magnetization vanishes, whereas at high
temperatures it vanishes as well because all quantum effects are
washed out by thermal fluctuations leading to random spin
orientations.

The striking contrast between the SQHD prediction,
Eq.~(\ref{eq:spin-current}) and Eq. (\ref{eq:s+s-}) is surprising
since the theoretical concepts that are used in condensed matter
physics include correlations and spin effects in a much more
accurate fashion than QHD.
As is pointed out in the analysis of Ref. \cite{spin_critique} the SQHD analysis contains a major inconsistency
(see our discussion above): the $N$-particle wave function is represented by a product of single-particle
orbitals (Hartree or Vlasov approximation) whereas for fermions an anti-symmetrized ansatz has to be used.
This leads to Slater determinants that guarantee the Pauli principle, in contrast to the Hartree ansatz. As a result
the Fermi statistics and the Pauli principle are lost in key places of the QHD theory. In particular, kinetic
effects such as the sharp Fermi surface are lost\footnote{The Fermi distribution is only included via the equation of
state relating pressure and density, but this introduces degeneracy effects only in an average fashion.}.

Therefore, the predictions of exotic spin quantum effects in quantum plasmas such as spin-gradient-driven light amplification
\cite{braun_prl12} have to be questioned. While there is always room for new fascinating discoveries they, obviously, have
to be based on a well established theory that includes all relevant effects. An urgent next step, to resolve these
conflicting predictions is, therefore, to reformulate QHD fully using anti-symmetric $N-$particle states, thus building
in the Pauli principle from the very beginning.

\section{Conclusion and Outlook}\label{s:conclusion}

In this chapter, we have discussed the theoretical treatment of
dense quantum plasmas that are increasingly important in many
laboratory and astrophysical systems. While accurate approaches to
quantum plasmas have been in existence for many years -- based on
first-principle simulation, quantum kinetic theory and
non-equilibrium Greens functions -- these approaches are quite
difficult, especially for magnetized plasmas. This makes it highly
desirable to have at hand simpler models. Here quantum hydrodynamics
has become quite popular in various scientific areas, as is evident
from the vast literature that appeared in recent years. At the same
time, most papers have essentially ignored the limited applicability
range of QHD raising questions about the reliability of the results
and of their relevance for practical applications.

In this chapter, we have discussed the main concepts of quantum
hydrodynamics and its relations to quantum kinetic theory in terms
of the Wigner distribution function and its equation of motion. We
analyzed in detail the basic assumptions that lead to the QHD
equations and their limitations. Strictly speaking, QHD applies to
an ideal  Fermi gas (where the quantum coupling parameters are
small, i.e. $r_s\ll 1$ and $\Gamma_q \ll 1$) at zero temperature,
and it entirely neglects quantum exchange effects. Furthermore, as
any hydrodynamic theory, QHD is only able to resolve processes at
sufficiently large length scales  exceeding a threshold which is on
the order of the Thomas-Fermi screening length $\lambda_{TF}$.
Furthermore, QHD uses a closure of the system of
hydrodynamic equations that involves an equation of state in the local
approximation which again rules out strong inhomogeneities.
When any of these inherent limitations is neglected, unphysical
results can follow which includes the predictions of attractive
forces between protons at atomic scales as well as giant
magnetizations related to the electron spin.

In the linearized QHD, we have briefly reviewed some properties of
electron and ion plasma oscillations for unmagnetized as well as
magnetized quantum plasma. These include the linear electron plasma
waves in the strong degeneracy limit at $T=0 $ as well as for finite
$T$ and their dispersion. This was compared with the results from
kinetic theory which also provide information on the damping of the
oscillations. We further considered the quantum streaming and
Buneman instabilities, and low frequency electrostatic and
electromagnetic ion modes in uniform and nonuniform quantum plasma.
Finally, an overview of the nonlinear solutions of the QHD equations
was given, leading to localized coherent structures and the model
equations for a magnetized plasma have been presented with
illustrations for the sake of generality.

\textbf{Acknowledgements}
The authors thank Tim Schoof for valuable remarks.
This work has been supported by the Deutsche Forschungsgemeinschaft via SFB-TR24.

\printindex

\begin{thebibliography}{99}
\bibitem{mb1} M. Bonitz, N. Horing, P. Ludwig (eds.), \textit{Introduction to Complex Plasmas} (Springer, Berlin, 2010) Chs. 3 \& 4

\bibitem{sm} S.A. Maier, \textit{Plasmonics-Fundamentals and Applications}
(Springer, New York, 2007)

\bibitem{ha} H.A. Atwater, A. Polman, Plasmonics for improved photovoltaic
devices, Nature Mat.\textbf{\ 9}, 205 (2010) and references therein

\bibitem{mb2} M. Bonitz, R. Binder, S.W. Koch, Carrier-acoustic plasmons
instability in semiconductor quantum wires, Phys. Rev. Lett.
\textbf{70}, 3788 (1993)

\bibitem{mb3} M. Bonitz, R. Binder, D.C. Scott, S.W. Koch, D. Kremp,
Theory of plasmons in quasi-one-dimensional degenerate plasmas,
Phys. Rev. E \textbf{49}, 5535 (1994)

\bibitem{sg1} S.H. Glenzer, R. Redmer, X-ray Thomson scattering in high
energy density plasmas, Rev. Mod. Phys. \textbf{81}, 1625 (2009)

\bibitem{cr} C.P. Ridgers, et al., Dense Electron-Positron Plasmas and
Ultraintense $\gamma$--rays from Laser-Irradiated Solids, Phys. Rev.
Lett. \textbf{108}, 165006 (2012)

\bibitem{chabrier} G. Chabrier, D. Saumon, A.Y. Potekhin, Dense plasmas in astrophysics: from giant planets to
neutron stars, J. Phys. A: Math. Gen. \textbf{39} 4411 (2006)

\bibitem{mt} M.H. Thoma, Strongly coupled plasma in high energy physics, IEEE
Trans. Plasma Sci, \textbf{32}, 738 (2004)

\bibitem{vt} V.I. Tatarskii, The Wigner representation of quantum
mechanics, Sov. Phys. Usp. \textbf{26}, 311 (1983) [Usp. Fis. Nauk.
\textbf{139}, 587 (1983)]

\bibitem{db1} D. Bohm, D. Pines, A Collective Description of Electron
Interactions: III. Coulomb Interactions in a Degenerate Electron
Gas, Phys. Rev. \textbf{92}, 609 (1953)

\bibitem{yk1} Yu.L. Klimontovich, and V.P. Silin, On the spectra of systems of interacting particles, (in Russian)
Zh. Eksp. Teor. Fiz. \textbf{23}, 151 (1952)

\bibitem{yk2} Yu.L. Klimontovich, V.P. Silin, The spectra of systems of
interacting particles, In: Plasma Physics ed. by J. Drummond
(McGraw-Hill, New York 1961)

\bibitem{dp1} D. Pines, A Collective Description of Electron Interactions:
IV. Electron Interaction in Metals. Phys. Rev. \textbf{92}, 626
(1953)

\bibitem{dp2} D. Pines, Quantum Plasma Physics: Classical and Quantum
Plasmas. J. Nucl. Energy, Part C\textbf{\ 2}, 5 (1961)

\bibitem{dp3} P. Nozi\`{e}res, D. Pines, \textit{The Theory of Quantum
Liquids} (Benjamin, New York, 1966)

\bibitem{gm} G.D. Mahan, \textit{Many-Particle Physics} 2nd edn. (Plenum,
New York, 1990)

\bibitem{mb4} M. Bonitz, D. Semkat (eds.), \textit{Introduction to
Computational Methods for Many-Body Physics} (Rinton, Princeton,
2006)

\bibitem{vf1} V.S. Filinov et al., Monte Carlo simulations of dense quantum
plasmas, J. Phys. A: Math. Gen. \textbf{39, }4421 (2006)

\bibitem{vf2} V.S. Filinov, M. Bonitz, A. Filinov, V.O. Golubnychiy, Wigner
Function Quantum Molecular Dynamics, In: \textit{Computational
Many-Particle Physics, Lec. Notes in Phys. 739,} ed. by H. Fehske,
R. Schneider, A. Wei$\beta$e (Springer, Berlin, 2008)

\bibitem{mb5} M. Bonitz, A. Filinov, V.O. Golubnychiy, Th. Bornath, W.D.
Kraeft, First Principle Thermodynamic and Dynamic Simulations for
Dense Quantum Plasmas, Contrib. Plasma Phys. \textbf{45}, 450 (2005)

\bibitem{ph} P. Hohenberg, W. Kohn, Inhomogenous Electron Gas, Phys. Rev.
\textbf{136}, B864 (1964)

\bibitem{ws} W. Kohn, L.J. Sham, Self-Consistent Equations Including Exchange
and Correlation Effects, Phys. Rev. \textbf{140}, A1133 (1965)

\bibitem{dharma} M.W.C. Dharma-wardana, The Classical-Map Hyper-Netted-Chain
(CHNC) Method and Associated Novel Density-Functional Techniques for
Warm Dense Matter, Int. J. Quantum Chemistry {\bf 112}, 53 (2012)

\bibitem{perrot-dharma} F. Perrot, M.W.C. Dharma-wardana, Spin-polarized electron
liquid at arbitrary temperatures: Exchange-correlation energies,
electron-distribution functions, and the static response functions,
Phys. Rev. B {\bf 62}, 16536 (2000)

\bibitem{dufty-dutta} J.W. Dufty, S. Dutta, Classical representation of a quantum system at equilibrium: Theory, Phys. Rev. E {\bf 87}, 032101 (2013)

\bibitem{dutta-dufty1} S. Dutta, J.W. Dufty, Classical representation of a quantum system at equilibrium:
Applications, Phys. Rev. E {\bf 87}, 032102 (2013)

\bibitem{dutta-dufty2} S. Dutta, J.W. Dufty, Uniform electron gas at warm, dense matter
conditions, Eur. Phys. Lett. {\bf 102}, 67005 (2013)

\bibitem{mb6} M. Bonitz, \textit{Quantum Kinetic Theory} (Teubner,
Stuttgart, 1998)

\bibitem{dk1} D. Kremp, Th. Bornath, M. Bonitz, M. Schlanges, Quantum
kinetic theory of plasmas in strong laser fields, Phys. Rev. E
\textbf{60}, 4725 (1999)

\bibitem{mb7} M. Bonitz, Th. Bornath, D. Kremp, M. Schlanges, W.D. Kraeft,
Quantum Kinetic Theory for Laser Plasmas. Dynamical Screening in
Strong Fields, Contrib. Plasma Phys. \textbf{39}, 329 (1999)

\bibitem{mb8} M. Bonitz, D. Kremp, D.C. Scott, R. Binder, W.D. Kraeft,
H.S. K\"ohler, Numerical analysis of non-Markovian effects in
charge-carrier scattering: one-time versus two-time kinetic
equations, J. Phys.: Condens. Matter \textbf{8}, 6057 (1996)

\bibitem{mb9} K. Balzer, M. Bonitz, \textit{Nonequilibrium Green's Function
Approach to Inhomogenous Systems, Lec. Notes in Phys. 867}
(Springer, Berlin, 2013)

\bibitem{kb} L.P. Kadanoff, G. Baym, \textit{Quantum Statistical Mechanics}
(W.A. Benjamin, New York 1962)

\bibitem{nk} N.-H. Kwong, M. Bonitz, Real-Time Kadanoff-Baym Approach to Plasma
Oscillations in a Correlated Electron Gas. Phys. Rev. Lett.
\textbf{84}, 1768 (2000)

\bibitem{dk2} D. Kremp, M. Schlanges, W.D. Kraeft, \textit{Quantum
Statistics of Nonideal Plasmas} (Springer, Berlin, 2005)

\bibitem{mb10} M. Bonitz et al., Theory and simulations of strong
correlations in quantum Coulomb systems, J. Phys. A: Math. Gen.
\textbf{36}, 5921 (2003)

\bibitem{mb11} M. Bonitz, J.W. Dufty, Quantum kinetic theory of metal clusters
in an intense electromagnetic field I, Cond. Matt. Phys. \textbf{7},
483 (2004)

\bibitem{mb12} M. Bonitz et al., Classical and quantum Coulomb
crystals, Phys. Plasmas \textbf{15}, 055704 (2008)

\bibitem{mb13} M. Bonitz, Kinetic theory for quantum plasmas, AIP Conf.
Proc. \textbf{1421}, 135 (2012)

\bibitem{si} S. Ichimaru, Strongly coupled plasmas: high-density classical
plasmas and degenerate electron liquids, Rev. Mod. Phys.
\textbf{54}, 1017 (1982)

\bibitem{ps} P.K. Shukla, B. Eliasson, Colloquium: Nonlinear collective
interactions in quantum plasmas with degenerate electron fluids,
Rev. Mod. Phys. \textbf{83}, 885 (2011)

\bibitem{sv} S.V. Vladimirov, Y.O. Tyshetskiy, On description of
collisionless quantum plasmas, Phys. Uspekhi \textbf{54}, 1243
(2011)

\bibitem{em} E. Madelung,  Quantum theory in hydrodynamic form (in German), 
Z. Physik \textbf{40}, 322 (1927)

\bibitem{gm1} G. Manfredi, How to model quantum plasmas, Fields Inst.
Commun. \textbf{46}, 263 (2005)

\bibitem{gm2} G. Manfredi, F. Haas, Self-consistent fluid model for a
quantum electron gas, Phys. Rev. B \textbf{64}, 075316 (2001)

\bibitem{gm3} G. Manfredi, P.-A. Hervieux, Y. Yin, N. Crouseilles, Collective
Electron Dynamics in Metallic and Semiconductor Nanostructures, In:
\textit{Atomic-Scale Modeling of Nanosystems and Nanostructured
Materials, Lec. Notes Phys. 795} (Springer, Berlin, 2010)

\bibitem{nc} N. Crouseilles, P.-A. Hervieux, G. Manfredi, Quantum hydrodynamic
Models for nonlinear electron dynamics in thin metal films, Phys.
Rev. B \textbf{78}, 155412 (2008)

\bibitem{fh1} F. Haas, \textit{Quantum Plasmas-An Hydrodynamic Approach}
(Springer, New York, 2011)

\bibitem{aj} A. J\"{u}ngel, \textit{Transport equations in Semiconductors} (Springer, Berlin, 2009)

\bibitem{cg} C.L. Gardner, The Quantum Hydrodynamic Model for Semiconductor
Devices, SIAM (Soc. Ind. Appl. Math.) J. Appl. Math. \textbf{54},
409 (1994)

\bibitem{ab} A. Bennett, Influence of the Electron Charge Distribution on
Surface-Plasmon Dispersion, Phys. Rev. B \textbf{1}, 203 (1970)

\bibitem{mm} M. Marklund, G. Brodin, L. Stenflo, C.S. Liu, New quantum limits
in plasmonic devices, Euro. Phys. Lett. \textbf{84,} 17006 (2008)

\bibitem{eg} E.P. Gross, Structure of a quantized vortex in boson system, Nuovo Cimento \textbf{20}, 454 (1961)

\bibitem{lp} L.P. Pitaevskii, Vortex Lines in an Imperfect Bose Gas, Zh. Eksp. Teor. Fiz. \textbf{40}, 646 (1961)
[Sov. Phys. JETP 13, 451 (1961)]

\bibitem{as} A. Serbeto, L.F. Monteiro, K.H. Tsui, J.T. Mendonca, Quantum plasma
fluid model for high-gain free-electron lasers, Plasma Phys.
Control. Fusion \textbf{51,} 124024 (2009)

\bibitem{ak} A. Kendl, P.K. Shukla, Drift wave turbulence in a dense
semi-classical magnetoplasma, Phys. Lett. A \textbf{375}, 3138
(2011)

\bibitem{db2} D. Bohm, A Suggested Interpretation of the Quantum Theory in
Terms of "Hidden" Variables. I, Phys. Rev. 85, \textbf{166} (1952)

\bibitem{db3} D. Bohm, A Suggested Interpretation of the Quantum Theory in
Terms of "Hidden" Variables. II, Phys. Rev. 85, \textbf{180} (1952)

\bibitem{db4} D. Bohm, J.P. Vigier, Model of the Causal Interpretation of
Quantum Theory in Terms of a Fluid with Irregular Fluctuations.
Phys. Rev. \textbf{96}, 208 (1954)

\bibitem{phol} P.R. Holland, \textit{The Quantum Theory of Motion} (Cambridge,
New York, 1993)

\bibitem{rw} R.E. Wyatt. \textit{Quantum Dynamics with Trajectories}
(Springer, Berlin, 2005)

\bibitem{ew} E.P. Wigner, On the Quantum Correction for Thermodynamic
Equilibrium, Phys. Rev. \textbf{40}, 749 (1932)

\bibitem{jm} J.E. Moyal, Quantum mechanics as a statistical theory, Proc.
Cambridge Phil. Soc. \textbf{45}, 99 (1949)

\bibitem{pc} P. Carruthers, F. Zachariasen, Quantum collision theory with
phase-space distributions, Rev. Mod. Phys. \textbf{55}, 245 (1983)

\bibitem{mh} M. Hillery, R.F. O'Connell, M.O. Scully, E.P. Wigner, Distribution
functions in physics: Fundamentals, Phys. Rep. \textbf{106}, 121
(1984)

\bibitem{hl} H.W. Lee, Theory and Application of the Quantum Phase-Space
Distribution-Functions, Phys. Rep., \textbf{259},147 (1995)

\bibitem{jd} J. Dawson, On Landau Damping, Phys. Fluid \textbf{4}, 869 (1961)

\bibitem{fh2} F. Haas, G. Manfredi, M. Feix, Multistream model for quantum
plasma, Phys. Rev. E \textbf{62}, 2763 (2000)

\bibitem{bonitz_comment} M. Bonitz, E. Pehlke, T. Schoof, Attractive forces between ions in quantum plasmas: Failure of linearized quantum hydrodynamics, Phys. Rev. E {\bf 87}, 033105 (2013)

\bibitem{jl} J. Lindhard, K. Dan. Vidensk. Selsk. Mat. Fys. Medd. \textbf{28}, 8 (1954)

\bibitem{vg} V.O. Golubnychiy, M. Bonitz, D. Kremp, M. Schlanges, Dynamical properties
of plasmon dispersion of a weakly degenerate correlated
one-component plasma, Phys. Rev. E \textbf{64}, 016409 (2001)

\bibitem{dm1} D.M Melrose, \textit{Quantum Plasmadynamics: Unmagnetized
plasmas, Lec. Notes in Phys 735} (Springer, New York, 2008)

\bibitem{dm2} D.B. Melrose, A. Mushtaq, Quantum recoil and Bohm
diffusion, Phys. Plasmas \textbf{16}, 094508 (2009)

\bibitem{bonitz_pop94} M. Bonitz, Impossibility of plasma instabilities in isotropic quantum plasmas, Phys. Plasmas {\bf 1}, 832 (1994)

\bibitem{wf} W.R. Frensley, Boundary conditions for open quantum systems
driven far from equilibrium, Rev. Mod. Phys. \textbf{62}, 745 (1990)

\bibitem{aa} A.F. Alexandrov, L.S. Bogdankevich, A.A. Rukhadze, Osnovy
Elektrodinamiki Plazmy (\textit{Principles of Plasma
Electrodynamics}) (Moscow: Vysshaya Shkola, 1988) [Translated into
English (Springer, Berlin, 1984)]

\bibitem{fh3} F. Haas, L.G. Garcia, J. Goedert, G. Manfredi, Quantum
ion-acoustic waves, Phys. Plasmas \textbf{10}, 3858 (2003)

\bibitem{fh4} F. Haas, A. Bret, Nonlinear low frequency collisional quantum
Buneman instability, Europhys. Lett. \textbf{97}, 26001 (2012)

\bibitem{buneman} O. Buneman, Dissipation of Currents in Ionized Media, Phys. Rev. {\bf {115}}, 503 (1959)

\bibitem{bauch_springer} S. Bauch et al., Introduction to Quantum Plasma Simulations, Chapter in: \textit{Introduction to Complex Plasmas}. M. Bonitz, N. Horing, and P. Ludwig (eds.),
(Springer, Berlin, 2010)

\bibitem{zakharov} V.E. Zakharov, Collapse of Langmuir waves, Sov. Phys. JETP \textbf{35}, 908
(1972)

\bibitem{lg} L.G. Garcia, F. Haas, L. de Oliveira, J. Goedert, Modified
Zakharov equations for plasmas with a quantum correction, Phys.
Plasmas \textbf{12}, 012302 (2005)

\bibitem{fh5} F. Haas, P.K. Shukla, Quantum and classical dynamics of Langmuir
wave packets, Phys. Rev. E \textbf{79}, 066402 (2009)

\bibitem{sk1} S.A. Khan, H. Saleem, Linear coupling of Alfven waves and
acoustic-type modes in dense quantum magnetoplasmas, Phys. Plasmas
\textbf{16}, 052109 (2009)

\bibitem{hs} H. Saleem, Ali Ahmad, S.A. Khan, Low frequency electrostatic and
electromagnetic modes of ultracold magnetized nonuniform dense
plasmas, Phys. Plasmas \textbf{15}, 094501 (2008)

\bibitem{bs} B. Shokri, A.A. Rukhadze, Quantum drift waves, Phys. Plasmas
\textbf{6}, 4467 (1999)

\bibitem{shukla_prl12} P.K. Shukla, B. Eliasson, Novel Attractive Force between Ions in Quantum
Plasmas, Phys. Rev. Lett. {\bf 108}, 165007 (2012); Erratum: Phys.
Rev. Lett. {\bf 108}, 219902 (2012); Erratum: Phys. Rev. Lett. {\bf
109}, 019901 (2012)

\bibitem{shukla_comment} P.K. Shukla, B. Eliasson, M. Akbari-Moghanjoughi, Comment on "Attractive forces between ions in quantum plasmas: Failure of linearized quantum hydrodynamics", Phys. Rev. E {\bf 87}, 037101 (2013)

\bibitem{bonitz_reply} M. Bonitz, E. Pehlke, T. Schoof, Reply to "Comment on Attractive forces between ions in quantum plasmas: Failure of linearized quantum hydrodynamics", Phys. Rev. E {\bf 87}, 037102 (2013)

\bibitem{bonitz_phys-scr13} M. Bonitz, E. Pehlke, T. Schoof, Comment on ``Discussion on ‘Novel attractive force between ions in quantum plasmas---failure of simulations based on a density functional approach’'',
accepted for publication in Phys. Scripta (2013); arXiv:1309.5897

\bibitem{marklund_prl07} M. Marklund, G. Brodin,
Dynamics of Spin-$\frac{1}{2}$ quantum plasmas, Phys. Rev. Lett.
{\bf 98}, 025001 (2007)

\bibitem{mahajan_prl11} S.M. Mahajan, F. Asenjo,
Vortical dynamics of spinning quantum plasma: helicity conservation,
Phys. Rev. Lett. {\bf 107}, 195003 (2011)

\bibitem{braun_prl12} S. Braun, F. Asenjo, S.M. Mahajan,
Spin-Gradient-Driven Light Amplification in a quantum plasma, Phys.
Rev. Lett. {\bf 109}, 175003 (2012)

\bibitem{spin_critique} G.S.~Krishnaswami, R. Nityananda, A. Sen, A. Tyagaraja,
A critique of recent theories of spin half quantum plasmas,
arxiv:1306.1774 (2013)

\bibitem{lali10} E.M. Lifshitz, L.P. Pitaevski, {\em Physical Kinetics} (Pergamon,
Oxford, 1981)

\bibitem{ashcroft-mermin} N.W. Ashcroft, N.D. Mermin, {\em Solid State Physics} (Harcourt Brace, New York,
1976)

\bibitem{kittel} C. Kittel, {\em Introduction to Solid State Physics} (John Wiley, New York,
1963)

\end{thebibliography}
\end{document}